\documentclass[11pt]{article}

\usepackage{graphicx} 
\usepackage{cite}     

\begin{document}

\def \d {{\rm d}}
\def \M{{\cal A'_+}}
\def \N{{\cal A'_\times}}
\def\St{{T}}
\def\Sx{{X}}
\def\Sy{{Y}}
\def\Sz{{Z}}
\def\G{G}
\def \A  {{\cal A}}
\def \pap{{_\parallel}}
\def \pa {{\partial}}

\font\vekt=csbxti10 scaled 1100
\def\vektEbar#1{\{\!{\,\hbox{\vekt e}^{_\parallel}}_{\!(#1)}(\tau)\}}
\def\vektE#1{\{\hbox{\vekt e}_{(#1)}(\tau)\}}
\def\vektebar#1{\!{\,\hbox{\vekt e}^{_\parallel}}_{\!(#1)}}
\def\vekte#1{\hbox{\vekt e}_{(#1)}}

\def\Z#1{Z^{(#1)}}
\def\bZ#1{\bar Z^{(#1)}}
\def\TTT#1{{\buildrel{.\mkern-1mu.\mkern-1mu.\mkern-1mu}\over#1}}     
\def\op#1{\mathord{\rm #1}}

\def\tdif#1{{\op{d}}#1}          
\def\tddif#1{\op{d}#1^2}         
\def\tdifdif#1{{\op{d}}^2#1}     
\def\tdr#1#2{\frc{\tdif{#1}}{\tdif{#2}}}                
\def\tdrdr#1#2{\frc{\tdifdif{#1}}{\tddif{#2}}}          

\def\ReseniA#1{\ifmmode{\rm A}^{^{_{\hskip-.3em(#1)}}}%
                 \else${\rm A}^{^{_{\hskip-.3em(#1)}}}\hskip-.3em$\fi}
\def\ReseniB#1{\ifmmode{\rm B}^{^{_{\hskip-.1em(#1)}}}%
                 \else${\rm B}^{^{_{\hskip-.1em(#1)}}}\hskip-.3em$\fi}
\def\ReseniC#1{\ifmmode{\rm C}^{^{_{\hskip-.05em(#1)}}}%
                 \else${\rm C}^{^{_{\hskip-.05em(#1)}}}\hskip-.3em$\fi}

\def\frc#1#2{%
  \mathchoice{%
    {\textstyle\strut#1\over\textstyle\strut#2}}%
    {{\textstyle\strut#1\over\textstyle\strut#2}}%
    {{\scriptstyle\strut#1\over\scriptstyle\strut#2}}%
    {{\strut#1\over\strut#2}}}

\newtoks\schemata \newdimen\tbs
\def\addschema#1{\schemata\expandafter{\the\schemata#1}}
\def\addseparator{\ifx\params\empty \else \addschema&\fi}
\def\vemznak#1#2;{\def\params{#2}\csname P:#1\endcsname}
\expandafter\def\csname P:c\endcsname{%
   \addschema{\hfil$##$\hfil}\addseparator}
\expandafter\def\csname P:r\endcsname{%
   \addschema{\hfil$##$}\addseparator}
\expandafter\def\csname P:l\endcsname{%
   \addschema{$##$\hfil}\addseparator}
\expandafter\def\csname P:|\endcsname{%
   \addschema{\vrule\hskip2\tbs}}
\def\genschemata{\ifx\params\empty
   \else\expandafter\vemznak\params;\expandafter\genschemata\fi}

\tbs=.2em
\def\\#1{\cr\noalign{\vskip#1}}
\def\pole#1#2{\vcenter{\offinterlineskip
   \schemata={\tabskip=2\tbs \strut}
  \def\params{#1}\genschemata \tabskip=\tbs
   \halign{\span\the\schemata \tabskip=\tbs \cr #2\crcr}}}
\def\matice#1#2{\left(\pole{#1}{#2}\right)}

\def\crA{\\{1mm}}
\def\eqref#1{(\ref{#1})}

\newcounter{citac}
\newcounter{mujcit}
\newcounter{tabulka}
\renewcommand{\thecitac}{\hbox{As\arabic{citac}}}

\title{Geodesic motion in the Kundt spacetimes and the character of envelope singularity}

\addvspace{1cm}

\author{
J. Podolsk\'y\thanks{E--mail: {\tt podolsky@mbox.troja.mff.cuni.cz}}
\  and
M. Bel\'a\v n\thanks{E--mail: {\tt halef@atrey.karlin.mff.cuni.cz}}
\\ Institute of Theoretical Physics, Charles University in Prague,\\
V Hole\v{s}ovi\v{c}k\'ach 2, 18000 Prague 8, Czech Republic.
}

\date{\today}

\maketitle
\begin{abstract}
We investigate geodesics  in specific Kundt  type~$N$ (or conformally flat)
solutions to Einstein's equations. Components of the curvature tensor in
parallelly transported tetrads are then explicitly evaluated and
analyzed. This elucidates some interesting global properties of the spacetimes, such
as an inherent rotation of the wave-propagation direction, or the character of singularities.
In particular, we demonstrate that the characteristic envelope singularity of the
rotated wave-fronts is a (non-scalar) curvature singularity, although all scalar invariants
of the Riemann tensor vanish there.


\end{abstract}

\section{Introduction}

In 1961 a general class of solutions to Einstein's equations which admit a
non-expanding, shear-free, and twist-free null congruence  was introduced and described
by Wolfgang Kundt  \cite{Kundt61,Kundt62,EK} (for recent reviews see, e.g.,
\cite{kramerbook}, \cite{PodOrt03}). This contained some  previously
known families, in particular the \emph{pp\,}-waves \cite{brinkmann,BalJef26,Brdicka51,BPR}
or the Nariai \cite{Nariai51} and Bertotti-Robinson \cite{LeviCivita17BR,Bertotti59,Robinson59}
universes. Within his large class, Kundt also identified new solutions of various types,
among them specific type~$N$ spacetimes representing plane-fronted gravitational waves which
were geometrically different from famous \emph{pp\,}-waves.
These so-called Kundt waves are either vacuum solutions or contain pure radiation.  Their
generalization to a non-vanishing value of the cosmological constant $\Lambda$ was
later found by Ozsv\'ath \emph{et al} \cite{OzsRobRoz85}
(see also \cite{GarPle81,Siklos85,VanGunNar90,BicPod99I}). Very recently,
these solutions have been further generalized in \cite{GDP04} where a complete family of Kundt waves
of type~III (which admit $\Lambda$ and/or pure radiation) has been derived and classified.
Geometry of the wave surfaces and their envelope singularity has also been analyzed.
For particular choices of the metric functions one recovers more special type~$N$
Kundt waves and conformally flat non-expanding spacetimes.

In fact, the family of conformally flat pure radiation Kundt metrics attracted attention
few years ago
\cite{Wils89,KoutMcInt96,EdgLud97l,EdgLud97,EdgLud97grg,Skea97,GriPod98,Barnes01}.
In general, they contain no invariants or Killing or homothetic vectors,
and thus provide an interesting exceptional case for invariant classification of exact solutions.
It was demonstrated by Skea \cite{Skea97} that, to distinguish the Wils
\cite{Wils89} metric within the more general Edgar--Ludwig solution \cite{EdgLud97l},
it is necessary to go as far as but no further than the fourth derivative of the
curvature tensor. This result supports the conjecture that any space-time can be uniquely
characterized using the derivatives of the curvature tensor in which no higher than
the fourth derivative is ever required.

Even more interestingly, it has been recently demonstrated
\cite{BicakPravda,Pravdovi} that for the above Kundt spacetimes {\em all curvature
invariants} of all orders {\em identically vanish}. They may thus play an important role in string theory and
quantum gravity since there are no quantum corrections to all perturbative orders \cite{Coley}.

Despite the fact that particular aspects of these spacetimes have been studied for decades, many
questions concerning their geometrical and physical properties still remain open.
For example, analysis of the geodesic deviation \cite{BicPod99II} revealed the
structure of relative motions of test particles in the Kundt gravitational waves but
it is not yet clear whether the geodesic motion itself does exhibit a chaotic behaviour,
analogous to the chaos in \emph{pp\,}-waves \cite{PPvlnychaos}.

We consider here vacuum Kundt type~$N$ waves but some results  also
apply to conformally flat pure radiation solutions of this non-expanding class.
The  purpose of our paper is to contribute to understanding of  their global
structure, namely the character of singularities.
For principal reasons this is not an easy task since
all curvature invariants --- which would unambiguously identify physical
singularities --- vanish. Thus, to analyse singularities in the Kundt spacetimes
(summarized here in section~\ref{solution}) it is necessary to study the behaviour
of free test particles, which is done in section~\ref{geodetikya}. In the subsequent
section~\ref{para} we find frames parallelly transported along timelike and null
geodesics, and in section~\ref{project} we project the curvature tensor onto these tetrads.
We employ these results for discussion of the singularities in the final section~\ref{chara}.

\section{The Kundt spacetimes}
\label{solution}
The metric of type~$N$ or conformally flat Kundt spacetimes,  which are either vacuum
or  contain pure radiation, can be written in the form (see, e.g., \cite{PodOrt03})
\begin{eqnarray}
&& \d s^2=2\,\d\zeta\,\d\bar\zeta-2\,Q^2\d u\,\d v+F\,\d  u^2\ ,
 \label{metric}\\
&& \qquad Q =  \zeta+\bar\zeta\ ,\quad  F =  2\,Q^2\,v^2-Q\,H\ ,\nonumber
\end{eqnarray}
where $H(\zeta,\bar\zeta,u)$ is a function of the spatial
coordinates $\zeta$, $\bar\zeta$, and of the retarded time $u$.
A~simple transformation $v=wQ^{-2}$ puts the metric into the Kundt canonical
form \cite{Kundt61,kramerbook}
 \begin{equation}
 \d s^2=2\,\d\zeta\,\d\bar\zeta-2\,\d u\,\left(\d w+W\,\d\zeta+\bar{W}\d\bar\zeta+{\cal H}\,\d u\right)\ ,
 \label{metricK}
\end{equation}
with ${W = -2wQ^{-1}}$, ${{\cal H} =  -w^2Q^{-2}+\frac{1}{2}Q\,H}$
(the \emph{pp\,}-waves corresponding to ${W=0}$ and ${{\cal H}=H}$ in
(\ref{metricK}) will not be investigated here). In the natural null tetrad
\begin{equation}
 {\bf k}  =  \pa_v\ , \quad
 {\bf l} = (F/2)Q^{-4}\,\pa_v+ Q^{-2}\,\pa_u\ ,\quad
 {\bf m} = \pa_{\bar{\zeta}}\ , \quad {\bf\bar m}  =  \pa_{\zeta}\ , \label{privtetrad}
\end{equation}
(${\bf k}$ is the quadruple principal null direction) the only non-vanishing
Weyl and Ricci scalars are
\begin{equation}
 \Psi_4    = \textstyle{\frac{1}{2}}\,Q^{-3}\,H_{,\zeta\zeta}\ , \qquad
 \Phi_{22} = \textstyle{\frac{1}{2}}\,Q^{-3}\,H_{,\zeta\bar\zeta} \ . \label{psi4phi22}
\end{equation}

The spacetimes are  thus \emph{conformally flat} when ${H_{,\zeta\zeta}=0}$, i.e.
if $H$ has a special form
\begin{equation}
 H_0  = A_0+A_1\zeta+\bar{A}_1\bar\zeta+A_2\zeta\bar\zeta\ ,\label{confflat}
\end{equation}
where $A_i$ are arbitrary complex functions of $u$.
These spacetimes  containing \emph{pure radiation} characterized by ${\Phi_{22}=A_2/(2Q^3)}$
were considered by Kundt \cite{Kundt61,Kundt62,EK} and others
\cite{Wils89,KoutMcInt96,EdgLud97l,EdgLud97,EdgLud97grg,Skea97,GriPod98,Barnes01}.
For $A_2=0$  the matter is absent, and $H_0$ can be set zero by
a suitable coordinate transformation \cite{BicPod99I}. Such solution with $H=0$
corresponds to flat Minkowski space, see transformation
(\ref{TransformationPKToMinkowski}) below.

\emph{Vacuum type~$N$} Kundt waves arise when ${\Phi_{22}=0}$, ${\Psi_4\not=0}$, i.e. for
\begin{equation}
 H = f(\zeta,u)+\bar{f}(\bar{\zeta},u)\ ,\label{vac}
\end{equation}
where $f$ is an arbitrary function of $\zeta$ and $u$, holomorphic in $\zeta$,
such that ${f_{,\zeta\zeta}\not=0}$.
These solutions (generalized also to admit a cosmological constant $\Lambda$) were
presented in \cite{OzsRobRoz85,GarPle81,BicPod99I,GDP04}.

For our  analysis of geodesics and singularities it is
convenient to rewrite the Kundt  solutions
in  real spatial coordinates $x$ and $y$, related to the complex coordinate $\zeta$ by
\begin{equation}
\zeta =(x+\hbox{i}\, y)/{\sqrt2}\ .\label{real}
\end{equation}
The metric (\ref{metric}) obviously takes the form
\begin{equation}
{\d s}^2={\d x}^2+{\d y}^2-4x^2\d u\,\d v
         +4(x^2v^2+x\G)\,{\d u}^2\ ,
\label{Metric}\end{equation}
where ${G=-H/(2\sqrt2)}$. For vacuum spacetimes the function $G(x,y,u)$
satisfies the Laplace equation, ${\G_{,xx}+\G_{,yy}=0}$.
The Christoffel symbols in  coordinates ${(v,x,y,u)\equiv(0,1,2,3)}$ are
\begin{eqnarray}
&&\Gamma^{0}_{01} =1/x\ ,\quad
  \Gamma^{0}_{03} =-2v\ ,\quad
  \Gamma^{0}_{13} =-(\G/x)_{,x}\ ,\quad
  \Gamma^{0}_{23} =-(\G/x)_{,y}\ ,\nonumber\\
&&\Gamma^{0}_{33} =4v^3+4v\G/x-(\G/x)_{,u}\ ,\quad
  \Gamma^{1}_{33} =-4xv^2-2\G-2x\G_{,x}\ ,\label{Christoffel}\\
&&\Gamma^{1}_{03} =2x\ ,\quad
  \Gamma^{2}_{33} =-2x\G_{,y}\ ,\quad
  \Gamma^{3}_{13} =1/x\ ,\quad
  \Gamma^{3}_{33} =2v\ ,\nonumber
\end{eqnarray}
and all independent non-vanishing components of the curvature tensor read
\begin{equation}
 R_{1313}=-2x\G_{,xx}\ ,\quad
 R_{1323}=-2x\G_{,xy}\ ,\quad
 R_{2323}=-2x\G_{,yy}\ .
\label{Riemann}
\end{equation}

\begin{figure}[t]
\centering
\includegraphics[height=4.8cm]{Valec}
\hskip2mm
\includegraphics[height=4.8cm]{Kuzel}
\hskip0mm
\includegraphics[height=4.8cm]{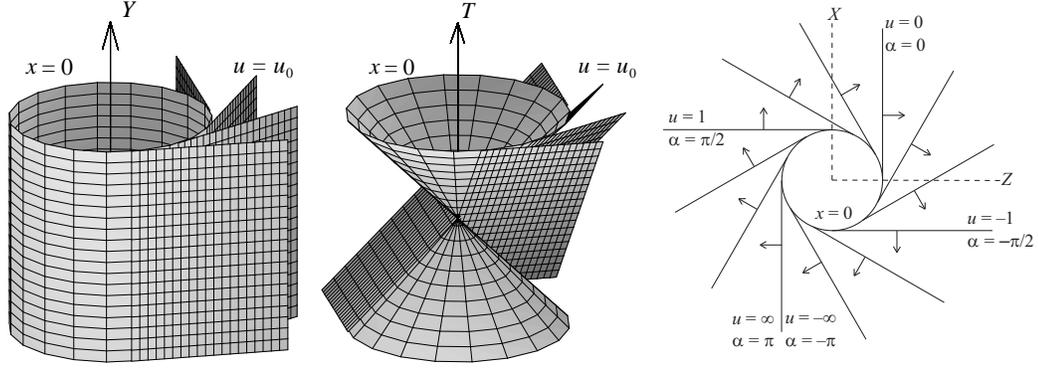}
\caption{\label{p1}
The wave surfaces  $u=u_0=\tan(\alpha/2)$ with $x\ge0$ are tangent
half-planes to expanding cylinder {(for ${\St=}$~const.; Left)}, cone {(for ${\Sy=}$~const.;
Middle)}, or  circle {(for ${\St,\Sy=}$~const.; Right)}
in Minkowski background coordinates. At any time $\St$ each wave surface is bounded
by the line  where it touches the cylinder ${x=0}$. For different values of $\,\alpha$, the
successive surfaces are rotated around the cylinder centered on the
$\Sy$-axis which expands with the speed of light
(four surfaces $u=u_0$ are indicated; Left and Middle). As
the circular caustic envelope ${x=0}$ expands (Right), the tangent half-planes
propagate in the spatial direction perpendicular to each wave surface.}
\end{figure}

Let us also recall the geometry of the Kundt spacetimes in the weak field
limit. It has been shown previously \cite{GriPod98} that the transformation
\begin{eqnarray}
\Sx&=&x\,(1+2uv)\ ,\qquad\qquad
\Sy=y \ ,\nonumber\\
\Sz&=&x\,[\,v-u(1+uv)\,]\ ,\ \quad
\St=x\,[\,v+u(1+uv)\,]\ ,\label{TransformationPKToMinkowski}
\end{eqnarray}
puts the Kundt metric \eqref{Metric} to
\begin{equation}
{\d s}^2={\d s}^2_0+
\frc{x\,G(x,y,u)}{{\Sx}^2+{\Sz}^2-{\St}^2}
        \Big[(1+u^2)\,\d{\St}-2u\,\d{\Sx}-(1-u^2)\,\d{\Sz}\Big]^2\  ,
        \label{MetricFlatPK}
\end{equation}
where
${\d s}^2_0=-{\d{\St}}^2+{\d{\Sx}}^2+{\d{\Sy}}^2+{\d{\Sz}}^2$
is the standard form of flat space corresponding to $\G=0$.
In (\ref{MetricFlatPK}),  $x$, $y$ and $u$ should be expressed using the
inverse of \eqref{TransformationPKToMinkowski}, namely
\begin{eqnarray} x&=&\pm\sqrt{{\Sx}^2+{\Sz}^2-{\St}^2}\ ,\ \qquad y=\Sy\ ,\nonumber\\
u&=&\frc{\Sx\mp\sqrt{{\Sx}^2+{\Sz}^2-{\St}^2}}{\St+\Sz}\ ,\quad
v=\pm\frc{\St+\Sz}{2\sqrt{{\Sx}^2+{\Sz}^2-{\St}^2}}\ .\label{TransformationMinkowskiToPK}
\end{eqnarray}

In the weak field limit (for small $G$) when the metric \eqref{MetricFlatPK}
is a perturbation of Minkowski background \cite{GriPod98}, each wave surface
${u=u_0=const.}$ corresponds to a hyperplane
\begin{equation}
-\St+\sin\alpha\,\,\Sx+\cos\alpha\,\,\Sz=0\ ,
\qquad \Sy \hbox{ arbitrary}\ ,\label{hyperplane}
\end{equation}
where \ $\alpha=2\arctan u_0$.
As $u_0$ increases from $-\infty$ to $+\infty$ then $\alpha$ goes from $-\pi$ to
$+\pi$, and these wave surfaces form a family of null hyperplanes
which rolls all around the cylinder
\begin{equation}
\Sx^2+\Sz^2=\St^2\ ,
\qquad \Sy \hbox{ arbitrary}\ .\label{cone}
\end{equation}
For $\St>0$ its radius is expanding at the speed of light.
In fact, it follows from \eqref{TransformationMinkowskiToPK} that this
caustic surface, formed as the envelope of all rotated wave surfaces,
corresponds to a singularity ${x=0}$ in the metric~\eqref{Metric},
see also~\cite{Urbantke}.

To obtain a unique foliation \eqref{hyperplane} of the region outside
the expanding cylinder  \eqref{cone} by the wave surfaces $u=u_0\in(-\infty,+\infty)$, it is necessary
to restrict these to half-hyperplanes. This is
achieved by considering only the range ${x\ge0}$ (or ${x\le0}$).
In both cases the wave surfaces are localized \emph{outside}
the cylinder \eqref{cone} given by ${x=0}$ which is formed as their envelope
as indicated in figure~\ref{p1}. The restriction ${x\ge0}$
also resolves the ambiguity in the signs of \eqref{TransformationMinkowskiToPK}
which corresponds to the invariance
of the  metric \eqref{Metric} under the reflections $x\to-x$, $G\to-G$.

Natural questions arise concerning the character of the envelope ${x=0}$
of the rotating family of wave fronts, and the  region inside it which is \emph{not covered}
by the Kundt coordinates. Is this \emph{envelope singularity} only
a coordinate one or is it a ``physical'' singularity?
Is it possible to extend geodesics  from the outside region across
${x=0}$? And what tidal forces would the corresponding geodesic observers feel?
We concentrate on these  open problems in our paper.

\section{Geodesics in the Kundt spacetimes}
\label{geodetikya}

Considering the Christoffel symbols \eqref{Christoffel} the geodesic
equations read
\begin{eqnarray}
\ddot{x}&=&-4x\dot{v}\dot{u}+(4xv^2+2\G+2x\G_{,x})\dot{u}^2\ ,\label{gx}\\
\ddot{y}&=&2x\G_{,y}\dot{u}^2\ ,\label{gy}\\
\ddot{u}&=&-2(\dot{x}/x)\,\dot{u}-2v\dot{u}^2\ ,\label{gu}
\end{eqnarray}
(where the dot denotes differentiation with respect to an affine
parameter $\tau$) plus a complicated equation for $\ddot{v}$.
Instead of it, however,  we take the normalisation condition of
four-velocity:
\begin{equation}
\dot{x}^2+\dot{y}^2-4x^2\dot{u}\dot{v}+4x^2\left({v^2+\G/x}\right)\dot{u}^2=\epsilon,
\label{Normalization}
\end{equation}
where $\epsilon$ denotes the character of the geodesic, namely
\begin{equation}
\epsilon=\cases{-1 & \cr
                 \hfill 0 & \cr
                +1 & \cr}
 \hbox{for }
         \cases{ \emph{timelike}  &\cr
                 \emph{null}      &\cr
                 \emph{spacelike} &\cr}
 \hbox{ geodesics. }
\label{DefinitionEpsilon}\end{equation}

Specific geodesic motion given by  equations \eqref{gx}-\eqref{Normalization}
depends on the particular spacetime function $\G(x,y,u)$. We
concentrate here on type~$N$ vacuum Kundt waves given by the
expression~\eqref{vac}. If the function $f(\zeta,u)$
is linear in $\zeta$, this generates only a flat space,
cf. \eqref{psi4phi22}. The simplest radiative
spacetimes thus arise when ${f=d(u)\zeta^n}$, ${n=2,3,4,\ldots}$. For a real
function $d(u)$, in coordinates $x,y$ this corresponds to
\begin{equation}
\G^{(n)}=c(u)\,\op{Re}\{(x+\hbox{i}\, y)^n\}\ .
\label{GObecna}
\end{equation}
An arbitrary function $c(u)$  determines the profile of
gravitational wave. Various sandwich and impulsive waves can thus
be constructed \cite{Podolsky98}. However, in our contribution we
only consider spacetimes for which the function
$\G$ is \emph{independent of $u$}, i.e. ${c(u)=const.}$ Without
loss of generality we assume ${c=1}$. Indeed,  employing the
scaling freedom of the metric \eqref{Metric},
${u\to\tilde{u}=\lambda\,u}$,\
${v\to\tilde{v}=\lambda^{-1}\,v}$,\
${\G\to\tilde{\G}=\lambda^{-2}\,\G}$,\
the factor $c$ can be set to unity by an appropriate choice of $\lambda$
(${\G<0}$ corresponding to ${c=-1}$ is related to ${\G>0}$
with ${c=1}$ by the reflection ${x\to-x}$).

\begin{figure}[t]
\centering
\includegraphics[height=3.2cm]{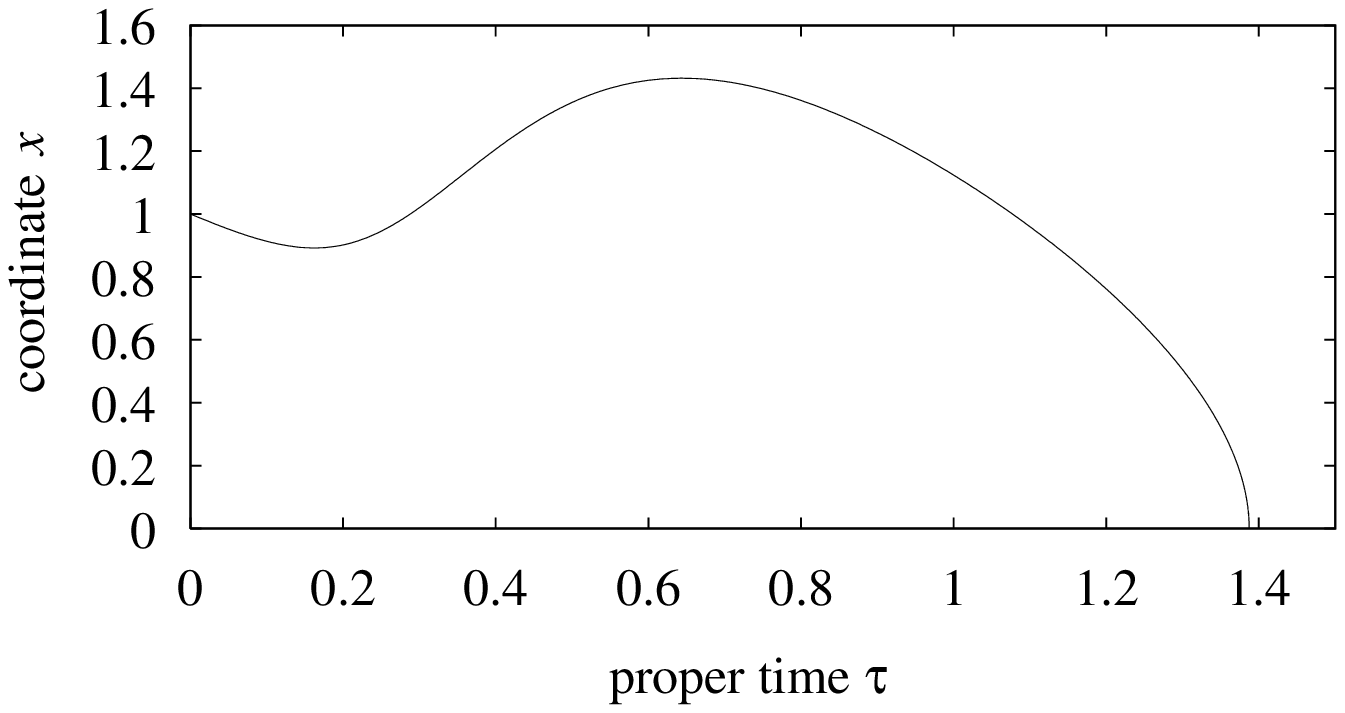}
\includegraphics[height=3.2cm]{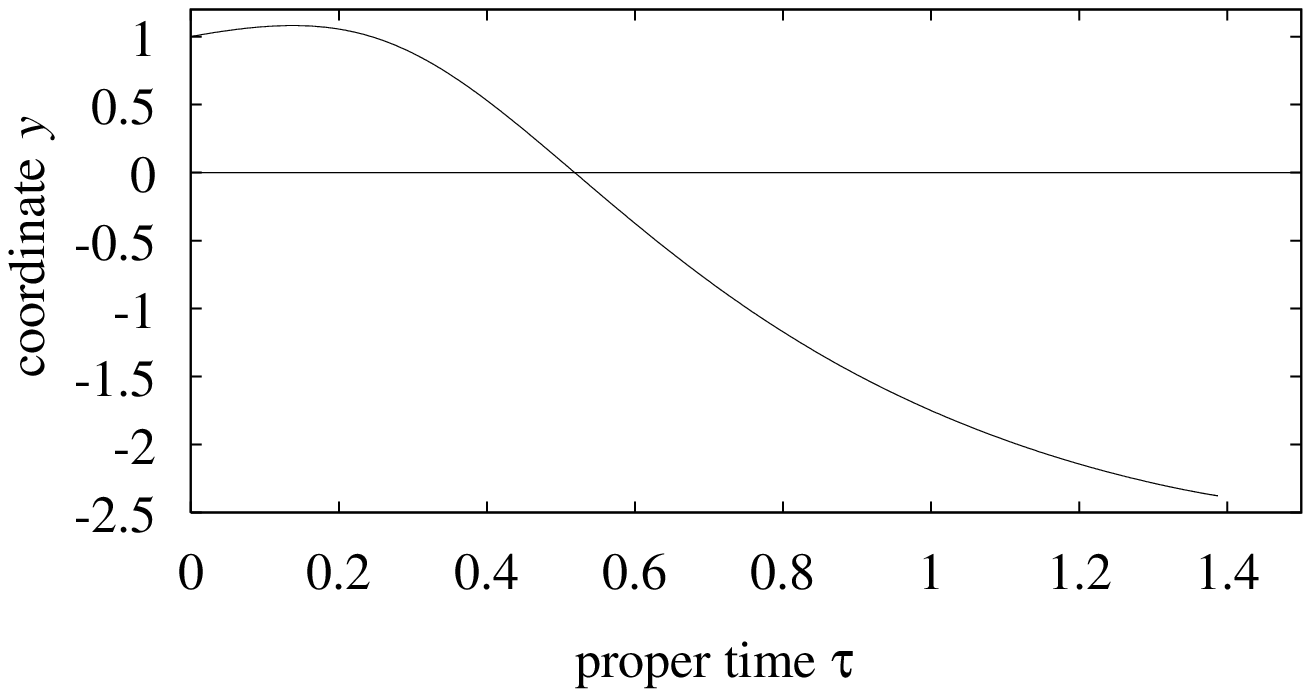}
\bigskip
\includegraphics[height=3.2cm]{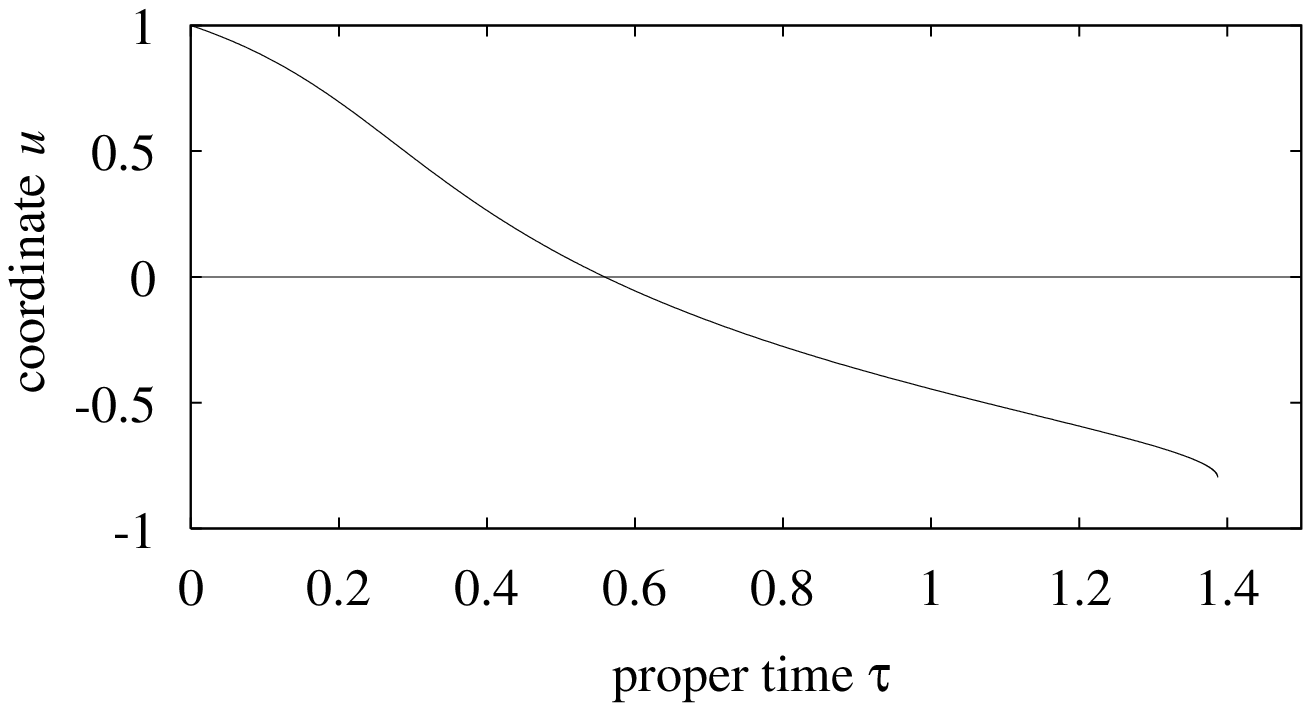}
\includegraphics[height=3.2cm]{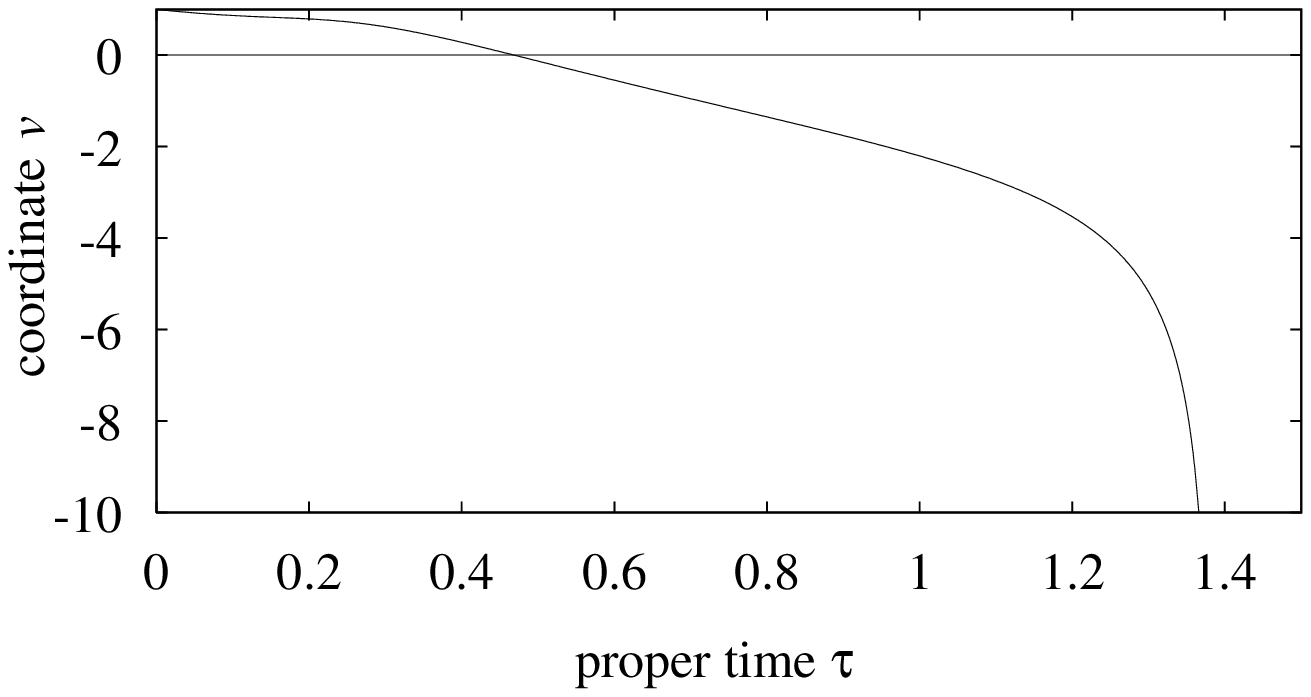}
\vspace*{-0.8cm}
\caption{\label{typicalg}
Typical behaviour of timelike geodesics in the Kundt wave spacetime ${n=2}$.
The initial values are $x=y=u=v=1$,
$\dot{x}=\dot{u}=-1$,
$\dot{y}=1$,
$\dot{v}=-2.75$ at ${\tau=0}$.
}
\end{figure}

Even after these simplifications are made, it is
difficult to find explicit solutions of equations
\eqref{gx}-\eqref{Normalization}. To
obtain more  general insight into the  geodesic motion  it is
necessary to perform numerical simulations. For example, typical behaviour
of a test particle in the Kundt spacetime with ${\G^{(2)}=x^2-y^2}$
is indicated in figure~\ref{typicalg}.
It reaches the envelope singularity ${x=0}$ in
a finite value of the proper time $\tau$, with the remaining coordinates decreasing.
However, for different initial values the coordinate
$x$ may grow to infinity. Discussion of our numerical
studies will be presented elsewhere
\cite{BelanPodolsky}. Instead, we concentrate here on derivation of
families of explicit geodesics which could be used for \emph{analytic} investigation of
the singularities.

\subsection{Explicit  geodesics in the hypersurface ${y=0}$}

\begin{figure}[ht]
\centering
\includegraphics[height=3.2cm]{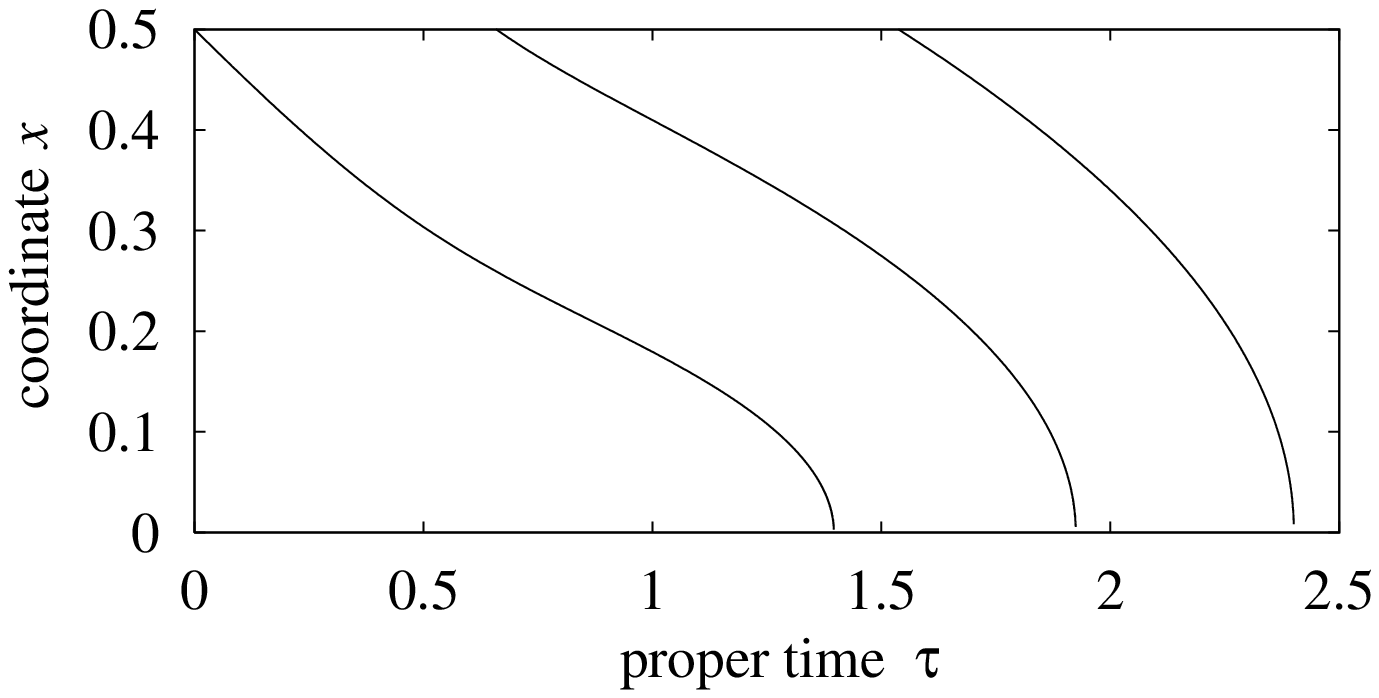}
\includegraphics[height=3.2cm]{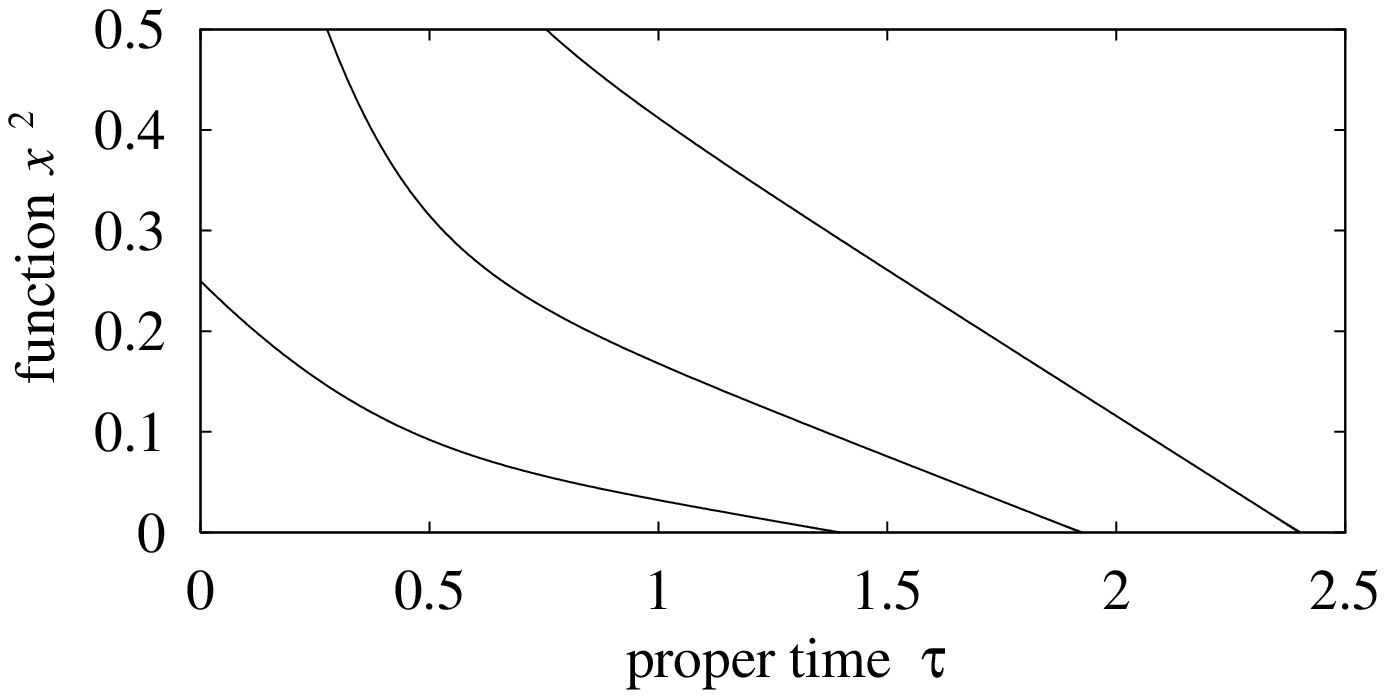}
\vspace*{-0.4cm}
\caption{\label{f3}
Typical null geodesics approaching the singularity at
 ${x=0}$ in the Kundt spacetime  ${n=2}$.
The initial values at ${\tau=0}$ are $x=0.5,1, 1.5$, and $\dot{u}=-0.45,-1.2,-2.17$,
respectively.
  }
\end{figure}

The equations simplify  if we
restrict to \emph{geodesics in the hypersurface} ${y=0}$. In fact, it
is a natural assumption: in the background Minkowski coordinates
this corresponds to ${\Sy=0}$ which is a plane perpendicular to the
symmetry axis of the cylindrical envelope singularity, see figure~\ref{p1}.
The expressions \eqref{GObecna} for
$\G^{(n)}$ (with ${c=1}$) and its derivatives on ${y=0}$  simplify to
\begin{equation}
\G^{(n)}=\>x^n\ ,\qquad
\G^{(n)}_{,x}=\>nx^{n-1}\ ,\qquad
\G^{(n)}_{,y}=\>0\ .\label{GKvadr-q}
\end{equation}
The equation \eqref{gy} is then identically satisfied by ${y(\tau)\equiv0}$.
Remaining equations \eqref{gx}, \eqref{Normalization},  \eqref{gu}  take the form
\begin{eqnarray}
\ddot{x}+4x\dot{v}\dot{u}-[4xv^2+2(n+1)x^n]\,\dot{u}^2&=&0\ ,\label{gxy=0}\\
\dot{x}^2-4x^2\dot{u}\dot{v}+4x^2(v^2+x^{n-1})\,\dot{u}^2&=&\epsilon\ ,\label{normy=0}\\
\ddot{u}+2(\dot{x}/x)\,\dot{u}+2v\dot{u}^2&=&0\ .\label{guy=0}
\end{eqnarray}
From the relation \eqref{guy=0} we easily express $v$ (assuming $\dot{u}\not=0$) and  substitute into
\eqref{gxy=0} and \eqref{normy=0}. We obtain a  system for the functions $\dot{u}$ and $x$ only,
\begin{eqnarray}
\label{GeodesicsHypersurface-1}
&&2\frc{\TTT{u}}{\dot{u}}-3\frc{\ddot{u}^2}{\dot{u}^2}+2(n+1)x^{n-1}\dot{u}^2+3\frc{\ddot{x}}{x}=0\ ,\\
\label{GeodesicsHypersurface-2}
&&2\frc{\TTT{u}}{\dot{u}}-3\frc{\ddot{u}^2}{\dot{u}^2}+4x^{n-1}\dot{u}^2+\frc{\dot{x}^2}{x^2}+4\frc{\ddot{x}}{x}=\frc{\epsilon}{x^2}\ .
\end{eqnarray}
Subtracting \eqref{GeodesicsHypersurface-2} from \eqref{GeodesicsHypersurface-1} we derive an important relation
\begin{equation}
{\left({\textstyle\frac{1}{2}}{x^2}\right)}^{\cdot\cdot}
\equiv x\ddot{x}+\dot{x}^2\  = \epsilon+2(n-1)x^{n+1}\dot{u}^2.
\label{GeodesicsHypersurface-U}
\end{equation}
It is possible to express $\dot{u}$ and substitute into \eqref{GeodesicsHypersurface-1}
but the resulting fourth-order non-linear equation for  $x$ is
very complicated, and additional assumptions are
thus necessary. The term on the right-hand side of \eqref{GeodesicsHypersurface-U}
indicates that the relation between $\dot{u}$ and powers of $x$ is
essential for the character of solution. For example, if ${x^{n+1}{\dot u}^2\to0}$
then it becomes
${{(x^2)}^{\cdot\cdot}\approx 2\epsilon}$, which implies
\begin{equation}
x\approx\sqrt{\epsilon\tau^2+C_1\tau+C_0}\ ,\label{xapr}
\end{equation}
where ${C_0, C_1}$ are constants. Such  geodesics
are possible: figure~\ref{f3} shows null geodesics
(${\epsilon=0}$) obtained by numerical simulations.
Linear behaviour of the function $x^2$ near ${x=0}$ is obvious.

To make the analysis more systematic let us \emph{assume a power-law relation}
between $\dot{u}$ and $x$,
\begin{equation}
\dot{u}=A\,x^a\ ,
\label{AnzatzUt}
\end{equation}
where ${a, A\not=0}$ are constants. Putting this into \eqref{GeodesicsHypersurface-U}
we get equation which can be integrated to
\begin{equation}
{\dot{x}}^2=\epsilon+\frac{4A^2(n-1)}{2a+n+3}\,x^{2a+n+1}+\frac{C}{x^2}\ ,
\label{AnzatzUt-1}
\end{equation}
with $C$ being the constant of integration. By substituting into \eqref{GeodesicsHypersurface-1}
we  derive the relation
\begin{equation}
-\epsilon\,T_1+4A^2 T_2\,x^{2a+n+1}+T_3\frac{C}{x^2}=0\ ,
\label{AnzatzUt-2}
\end{equation}
where the constant parameters are
\begin{eqnarray}
T_1&=&a(a+2)\ ,\nonumber\\
T_2&=&\frac{a+n+1}{2a+n+3}\Bigl[a(n-1)+2n\Bigr]\ ,\label{AnzatzUt-DefinitionT}\\
T_3&=&1-(2+a)^2\ .\nonumber
\end{eqnarray}
The equation \eqref{AnzatzUt-2} is \emph{exactly} satisfied if
$\epsilon=0$, $C=0$, $T_2=0$,  or if $\epsilon=\pm1$, $C=0$, $2a+n+1=0$,
i.e. for three possible choices of the parameter $a$,
\begin{eqnarray}
\hbox{(A)}\qquad a&=&-(n+1)\ ,\label{AnzatzUt-ExactChoicesA}\\
\hbox{(B)}\qquad a&=&-2n/(n-1)\ ,\label{AnzatzUt-ExactChoicesB}\\
\hbox{(C)}\qquad a&=&-(n+1)/2\ .\label{AnzatzUt-ExactChoicesC}
\end{eqnarray}
These cases enable three families of exact geodesics, denoted as
\ReseniA{n}, \ReseniB{n} and \ReseniC{n}, which will be discussed in detail in section \ref{ChapExact}.
Before this, however, we first investigate a general asymptotic behaviour
of solutions to \eqref{AnzatzUt-1}, \eqref{AnzatzUt-2}
as ${x\rightarrow0}$ or ${x\to\infty}$.

\subsection{Asymptotic behaviour of geodesics}
\label{ChapAnzatz-Discussion}
We start with  asymptotic solutions  as ${x\to0}$.
To avoid the divergence in the term $x^{-2}$ in \eqref{AnzatzUt-2} it is necessary to set either ${C=0}$ or ${T_3=0}$. All possibilities, summarized in table~\ref{1Table}, are:

\begin{list}{\thecitac.}
{\labelsep5pt\labelwidth30pt\topsep3pt\parsep3pt\itemsep0pt\setlength{\leftmargin}{35pt}\setlength{\rightmargin}{26pt}}
\refstepcounter{citac}
\label{TableNula-1}
\item The region $x\rightarrow0$ is reached  when ${C=0}$, ${\epsilon=0}$, and
$2a+n+1\equiv k>0$. These conditions imply ${x^{n+1}{\dot u}^2}\to0$,
i.e. near ${x=0}$ the solution approaches~$\eqref{xapr}$.%
\refstepcounter{citac}
\label{TableNula-4}
\item For ${a=-2}$ we obtain ${T_1=0}$. The region  ${x\to0}$ can only be approached asymptotically when
${C=0}$, ${n>3}$, ${\epsilon=0}$, and the right-hand side of \eqref{AnzatzUt-1} is positive.
This is a particular solution corresponding to the choice
${k=n-3}$ in \ref{TableNula-1}.%
\refstepcounter{citac}
\label{TableNula-2}
\item The choice ${a=-1}$ implies ${T_3=0}$, and only for ${\epsilon=0}$  the equation \eqref{AnzatzUt-2}
is fulfilled as ${x\to0}$. The right-hand side of equation \eqref{AnzatzUt-1} is then positive and
it can be solved if ${C>0}$. This asymptotic solution exactly corresponds to
\eqref{xapr}.%
\refstepcounter{citac}
\label{TableNula-3}
\item We can also make ${T_3=0}$ by choosing ${a=-3}$. Remaining conditions are the same as in
the case \ref{TableNula-2}, but we have to assume ${n>5}$ to reach ${x\to 0}$.
In contrast to \ref{TableNula-2}, ${T_2=2-n<0}$. This solution also corresponds  to \eqref{xapr}
asymptotically.
\end{list}
\begin{table}[h]
 \[
  \begin{array}{|c||c|c|c|c|} \hline
    & C &  a &  \hbox{Eq.~\eqref{AnzatzUt-2}}& \hbox{Eq.~\eqref{AnzatzUt-1}}
                          \\ \hline\hline
\ref{TableNula-1}&0&\,\frac{1}{2}(k-n-1)
              &x^{k}\sim0,k>0
              &\,{\dot{x}}^2=4A^2\frac{n-1}{k+2}x^{k}\\
  \hline
\ref{TableNula-4}&0&-2
              &x^{n-3}\sim0,n>3
              &{\dot{x}}^2=4A^2x^{n-3}\\
  \hline
\ref{TableNula-2}&>0&-1
              &x^{n-1}\sim0
              &{\dot{x}}^2\approx Cx^{-2} \\
  \hline
\ref{TableNula-3}&>0&-3
             &\,x^{n-5}\sim0,n>5
             &{\dot{x}}^2\approx Cx^{-2} \\
  \hline
 \end{array}
 \]
 \caption{Asymptotic solutions  as ${x\to0}$, assuming  ${\dot{u}=Ax^a}$. All
these geodesics are null (${\epsilon=0}$).}
\label{1Table}
\end{table}

Explicit forms of ${x(\tau)}$ corresponding to the above four
possibilities are summarized in table~\ref{2Table}.
Notice that the solutions \ref{TableNula-1}
with ${0<k<2}$ reach ${x=0}$ in the finite value $\tau_0$ of the affine  parameter,
whereas for ${k>2}$ they approach this singularity as ${\tau\to\infty}$.
The solution \ref{TableNula-4} exhibits an analogous behaviour, according to whether ${3<n<5}$ or ${n>5}$.
\begin{table}[h]
 \[
  \begin{array}{|c||c|} \hline
    & \hbox{explicit solutions as ${x\to0}$}     \\ \hline\hline
\ref{TableNula-1}
    &\,x\approx\left[|2-k|\sqrt{\frac{n-1}{k+2}}A\,(\tau-\tau_0)\right]^\frac{2}{2-k}\\
  \hline
\ref{TableNula-4}
    &x\approx\left[|5-n|A\,(\tau-\tau_0)\right]^\frac{2}{5-n}\\
  \hline
\ref{TableNula-2}
    & x\approx\left[2\sqrt{C}\,(\tau-\tau_0)\right]^\frac{1}{2}\\
  \hline
\ref{TableNula-3}
    & x\approx\left[2\sqrt{C}\,(\tau-\tau_0)\right ]^\frac{1}{2}\\
  \hline
 \end{array}
 \]
 \caption{Explicit asymptotic forms of ${x(\tau)}$ for null geodesics as ${x\to0}$.}
 \label{2Table}
\end{table}

We similarly investigate asymptotic geodesics in the region
${x\to\infty}$. It is not now in general necessary to set ${C=0}$ since the term  $Cx^{-2}$ in
\eqref{AnzatzUt-1}, \eqref{AnzatzUt-2} becomes negligible for large values of $x$.
Specific possibilities, summarised in table \ref{3Table}, are:
\begin{list}{\thecitac.}{\labelsep5pt\labelwidth30pt\topsep3pt\parsep3pt\itemsep0pt\setlength{\leftmargin}{35pt}\setlength{\rightmargin}{26pt}}
\refstepcounter{citac}
\label{TableNekonecno-1}
\item The region ${x\to\infty}$ is reached asymptotically if ${2a+n+1\equiv k<0}$ and ${\epsilon=0}$.
The case  ${k=-2}$ is, however, forbidden.
\refstepcounter{citac}
\label{TableNekonecno-2}
\item The choice ${a=-2}$ implies ${T_1=0}$, which requires ${n=2}$  and ${\epsilon=0}$
to reach the area ${x\rightarrow\infty}$. This is a particular case of \ref{TableNekonecno-1} for ${k=-1}$.
\refstepcounter{citac}
\label{TableNekonecno-3}
\item If ${a=-3}$ then ${T_3=0}$. Asymptotic solutions ${x\rightarrow\infty}$
now require  ${n=4}$ with ${\epsilon=0}$. The alternative case ${n=2}$ corresponds to
\ref{TableNekonecno-1} for ${k=-3}$.
\refstepcounter{citac}
\label{TableNekonecno-A}
\item The choice (A) given by (\ref{AnzatzUt-ExactChoicesA}) implies ${T_2=0}$. We thus have to put ${\epsilon=0}$.
\refstepcounter{citac}
\label{TableNekonecno-B}
\item Choosing (B), i.e. (\ref{AnzatzUt-ExactChoicesB}), we also obtain ${T_2=0}$ which implies ${\epsilon=0}$,
 ${n\not=3}$.
\refstepcounter{citac}
\label{TableNekonecno-C}
\item Finally, the class (C) given by (\ref{AnzatzUt-ExactChoicesC}) (corresponding to ${k=0}$)
is the only one which admits ${\epsilon=\pm1}$. The Eq.~(\ref{AnzatzUt-2}) requires
${2A^2=-\epsilon(n-3)/(n^2-4n-1)}$ which excludes ${n=3}$.
The case ${n=2}$ is inconsistent with Eq.~\eqref{AnzatzUt-1}, unless
${\epsilon=+1}$, ${x<0}$. For ${n=5,6,\ldots}$ the geodesics are timelike, for
${n=4}$ it is a spacelike geodesic.
\end{list}

\begin{table}[h]
 \[
  \begin{array}{|c||c|c|c|} \hline
    & a &  \hbox{Eq.~\eqref{AnzatzUt-2}}& \hbox{Eq.~\eqref{AnzatzUt-1}}
                          \\ \hline\hline
\ref{TableNekonecno-1}&\,\frac{1}{2}(k-n-1)
     &\,4A^2T_2x^k+CT_3x^{-2}\approx0,k<0
     &{\dot{x}}^2=4A^2\frac{n-1}{k+2}\,x^{k}+Cx^{-2}\\
  \hline
\ref{TableNekonecno-2}&-2
     &1/x\sim0 \hbox{\quad for\quad } n=2
     &{\dot{x}}^2\approx 4A^2/x  \\
  \hline
\ref{TableNekonecno-3}&-3
     &1/x\sim0 \hbox{\quad for\quad } n=4
     &{\dot{x}}^2\approx 12A^2/x \\
  \hline
\ref{TableNekonecno-A}&-(n+1)
     &Cx^{-2}\sim0
     &{\dot{x}}^2=-4A^2x^{-(n+1)}+Cx^{-2}\\
  \hline
\ref{TableNekonecno-B}&-2n/(n-1)
     &Cx^{-2}\sim0\ ,\ n\neq3
     &\,{\dot{x}}^2=\frac{4A^2(n-1)^2}{(n+1)(n-3)}x^{\frac{n^2-4n-1}{n-1}}+Cx^{-2}\\
  \hline
\mkern-2mu\ref{TableNekonecno-C}\mkern-2mu&-\frac{1}{2}(n+1)
     & Cx^{-2}\sim0\ ,\ n\neq3
     &{\dot{x}}^2=\frac{-4\epsilon}{n^2-4n-1}+Cx^{-2}\\
  \hline
 \end{array}
 \]
 \caption{Asymptotic solutions of the form $\dot{u}=Ax^a$ as $x\rightarrow\infty$ admit ${C\not=0}$.
 All geodesics are null (${\epsilon=0}$), except in the last case \ref{TableNekonecno-C} for which
 ${\epsilon\not=0}$.}
\label{3Table}
\end{table}

The corresponding asymptotic forms of ${x(\tau)}$ near ${x=\infty}$ are summarized in  table~\ref{4Table}.

\begin{table}[h]
 \[
  \begin{array}{|c||c|} \hline
    & \hbox{explicit solutions as ${x\to\infty}$}     \\ \hline\hline
\ref{TableNekonecno-1}
    & \ x\approx\left[(2-k)\sqrt{\frac{n-1}{k+2}}A\,(\tau-\tau_0)\right]^\frac{2}{2-k} \hbox{\ for\ } -2<k<0\ \\
    & \ x\approx\left[2\sqrt{C}\,(\tau-\tau_0)\right ]^\frac{1}{2} \hskip28mm\hbox{\ for\ \ } k<-2\ \\
  \hline
\ref{TableNekonecno-2}
    & x\approx\left[3A\,(\tau-\tau_0)\right ]^\frac{2}{3} \\
  \hline
\ref{TableNekonecno-3}
    & x\approx\left[3\sqrt{3}A\,(\tau-\tau_0)\right ]^\frac{2}{3} \\
  \hline
\ref{TableNekonecno-A}
    & -x\approx\left[(n+3)A\,(\tau-\tau_0)\right]^\frac{2}{n+3} \ \hbox{\ for\ } C=0\     \\
    &\ x\approx \left[2\sqrt{C}\,(\tau-\tau_0)\right ]^\frac{1}{2}\hskip11mm \hbox{\ for\ } C\not=0\ \\
  \hline
\ref{TableNekonecno-B}
    & x\approx \left[-A\frac{n^2-6n+1}{\sqrt{(n+1)(n-3)}}\,(\tau-\tau_0)\right ]^\frac{2(1-n)}{n^2-6n+1}
     \ \hbox{\ for\ } n>3\ \\
    &x\approx \left[2\sqrt{C}\,(\tau-\tau_0)\right ]^\frac{1}{2}\hskip17mm\ \hbox{\ for\ } n=2, C\not=0\ \\
  \hline
\ref{TableNekonecno-C}
    &      x\approx \sqrt{\frac{-4\epsilon}{n^2-4n-1}}\,(\tau-\tau_0) \\
  \hline
 \end{array}
 \]
 \caption{Explicit asymptotic  geodesics as ${x(\tau)\to\infty}$.}
 \label{4Table}
\end{table}

Let us  note that there exist some special \emph{non-trivial exact solutions}.
For example, the particular case $n=2$ of \ref{TableNekonecno-A} implies ${T_3=0}$, see
(\ref{AnzatzUt-DefinitionT}), and the normalization condition \eqref{AnzatzUt-2} is thus exactly
satisfied. The corresponding \emph{null} geodesics are  given by
\begin{equation}
\sqrt{Cx^2-4A^2x}\bigg({\frc{x}{2C}+\frc{3A^2}{C^2}}\bigg)
 +\frc{6A^4}{C^{5/2}}\ln{\bigg({\sqrt{Cx^2-4A^2x}+x\sqrt{C}-\frc{2A^2}{\sqrt{C}}}\bigg)}
 =\tau-\tau_0\ ,
\label{HypersurfaceNullExactSolution}\end{equation}
where  ${C>{4A^2}/{x}>0}$. For large $x$ this approaches the asymptotic form presented in table \ref{4Table}.

Similarly, the case $n=5$ of \ref{TableNekonecno-C} gives  \emph{timelike} geodesics (${\epsilon=-1}$) with $C>0$,
\begin{equation}
x=\sqrt{\left({\tau-\tau_0}\right)^2\pm2\sqrt{C}\left({\tau-\tau_0}\right)}\ .
\label{HypersurfaceEpsilonExactSolution}\end{equation}
This resembles the asymptotic solution  (\ref{xapr}) but there is a different sign of the  term ${(\tau-\tau_0)^2}$.

It follows from tables~\ref{1Table}-\ref{4Table} that the assumption \eqref{AnzatzUt} admits
a large class of asymptotic null geodesics on the hypersurface ${y=0}$, in particular in
the region ${x\to\infty}$ where the family is parametrized by four constants of integration,
namely ${k, A, C, \tau_0,}$ and $u_0$. The parameter $a$ is in general \emph{negative} (except
for the possibility \ref{TableNula-1}) so that ${\dot{u}\to\infty}$ as ${x\to0}$, and
${\dot{u}\to0}$ in the region ${x\to\infty}$. Therefore, ${u\to u_0}$ as ${x\to\infty}$
which means that the \emph{geodesics become asymptotically tangent} to the corresponding
wave surface.
Numerical simulations also indicate that near ${x\to 0}$ the possibility \ref{TableNula-2}
corresponding to ${a=-1}$ (i.e. ${\dot{u}\sim 1/x}$) is  preferred.

\subsection{Exact ``power-law" geodesics}\label{ChapExact}
We have already mentioned that there exist three explicit families  \ReseniA{n}, \ReseniB{n}, \ReseniC{n}
of {\it exact} solutions to the geodesic equations \eqref{gxy=0}-\eqref{guy=0} with $\G$ given by \eqref{GKvadr-q}.
These are particular subclasses ${C=0}$ of \ref{TableNekonecno-A}, \ref{TableNekonecno-B},
\ref{TableNekonecno-C}, respectively, cf. (\ref{AnzatzUt-ExactChoicesA})-(\ref{AnzatzUt-ExactChoicesC})
and table~\ref{3Table}. In view of  Eq. (\ref{AnzatzUt-1}),
the coordinate $x$  must be some power of the affine parameter $\tau$. It then follows
from Eqs. (\ref{AnzatzUt}) and (\ref{guy=0}) that  also the
coordinates $u$ and $v$ must have the same behaviour, i.e.
\begin{equation}
x=\alpha(\tau-\tau_0)^p\ , \quad
u=\beta (\tau-\tau_0)^q+u_0\ , \quad
v=\gamma(\tau-\tau_0)^r\ .
\label{ExactSolution}
\end{equation}
The constants ${\alpha, \beta, \gamma, p, q, r, \tau_0, u_0}$, related to  previous parameters by
\begin{equation}
a=(q-1)/p\ ,\qquad
A=q\,\beta/\alpha^a\ ,
\label{ExactSolutionConstants}\end{equation}
are restricted by the above equations. A straightforward calculation shows that
there are three distinct possibilities, according to the value of the parameter ${B\equiv\beta\gamma}$:
\begin{description}
\item[\ReseniA{n}\ ] These solutions which exist for all powers $n$ are given by  ${B=-1}$,
but ${\alpha^{n-1}<0}$ so that this only apply to even $n$
(implying $x<0$). The geodesics \ReseniA{n} are null and correspond to  \ref{TableNekonecno-A},
characterized by ${a=-(n+1)}$, with the choice ${C=0}$.%
\item[\ReseniB{n}\ ] The solutions \ReseniB{n} arise when ${B=-2/(n-1)^2}$ and
correspond to the ${C=0}$ subcase of \ref{TableNekonecno-B}, given by ${a=-2n/(n-1)}$. These
geodesics are also null in the region ${x<0}$ for $n=2$, or in the region ${x>0}$ if ${n>3}$.%
\item[\ReseniC{n}\ ] The solutions \ReseniC{n} for ${B=-\frac{1}{2}(n-3)/(n-1)}$ in
the same domains of $x$ as \ReseniB{n} are spacelike (${\epsilon=1}$) for ${n=2}$ or $4$,
and timelike (${\epsilon=-1}$) for $n>4$. This is the possibility \ref{TableNekonecno-C}, characterized by
${a=-(n+1)/2}$, for  $C=0$. The parameter $\gamma$ is now uniquely given as
${2\gamma^2=(n-3)\alpha^{n-1}}$. For ${n=2}$ this implies ${\alpha<0}$, i.e. ${x<0}$.
\end{description}
\begin{table}[h]
 \[
  \begin{array}{|c||c|c|c|c|c|c|} \hline
  & B & \epsilon & p& q=-r& \alpha^{n-1}&\beta   \\ \hline\hline
\hbox{\ReseniA{n}}
     &-1\
     &0
     &\frc{2}{n+3}
     & -\frc{n-1}{n+3}
     & -\frc{\gamma^2}{(n-1)^2}
     &-\frc{1}{\gamma}\\
  \hline
\hbox{\ReseniB{n}}
     &\frc{-2}{(n-1)^2}
     &0
     &\frc{2(1-n)\ }{n^2-6n+1}
     & \frc{(n-1)^2}{n^2-6n+1}
     &\frc{\gamma^2}{4}(n-3)(n+1)
     &\frc{-2}{\gamma(n-1)^2}\\
  \hline
\hbox{\ReseniC{n}}
     &\frc{-(n-3)}{2(n-1)}
     &\pm1
     &1
     &-\frc{n-1}{2}
     &\left(\frc{-4\epsilon}{n^2-4n-1}\right)^\frac{n-1}{2}
     &\frc{-1}{2\gamma}\frc{n-3}{n-1}\\
  \hline
 \end{array}
 \]
 \caption{Three classes of exact solutions in the hypersurface ${y=0}$.
 The families \ReseniA{n}, \ReseniB{n} of null geodesics are parameterised by an
 arbitrary constant $\gamma$.}
\label{Table-ExactSolutions1}
\end{table}

Finally, we also attempted to find  more general geodesics such that ${y\neq 0}$.
Assuming
\begin{equation}
x=\alpha(\tau-\tau_0)^p\ , \quad
u=\beta (\tau-\tau_0)^q+u_0\ , \quad
v=\gamma(\tau-\tau_0)^r\ , \quad
y=\delta(\tau-\tau_0)^s\ ,
\label{ExactSolution-y}\end{equation}
we found that there are no such geodesics when ${n>2}$.
However,  for  ${n=2}$ (i.e. ${\G^{(2)}=x^2-y^2}$) there exists a class
of exact null geodesics
\begin{equation}
x=\frac{3}{2}\gamma^2(\tau-\tau_0)^\frac{2}{5}\ ,\
v=\gamma(\tau-\tau_0)^\frac{1}{5}\ ,\
u=-\frac{1}{\gamma}(\tau-\tau_0)^{-\frac{1}{5}}+u_0\ ,\
y=\pm \frac{3\sqrt{5}}{2}\gamma^2(\tau-\tau_0)^\frac{2}{5}\ .
\label{ExactSolution-21}\end{equation}

\section{Parallelly transported frames}
\label{para}

After analyzing privileged families of geodesics in the Kundt spacetimes we proceed
to the study of singularities at ${x=0}$ and  ${x=\infty}$. Their character is not
obvious: they \emph{cannot be scalar curvature} singularities \cite{BicakPravda,Pravdovi}.
However, this leaves open the possibility that these are either non-scalar curvature
singularities or quasi-regular singularities, according to the classification scheme
introduced in \cite{EllisSchmidt}. To resolve this problem it is necessary to find
{\em orthonormal tetrads which are parallelly transported} along geodesics approaching
${x=0}$ and  ${x=\infty}$. Projections of the curvature tensor onto these
frames will then elucidate the character of the singularities.

\subsection{Frames parallelly transported along any timelike geodesic}
A natural ``interpretation" orthonormal frame $\vektE{i}$ for the Kundt spacetime
(\ref{Metric}),
\begin{eqnarray}
e^\mu_{(0)}&=&u^\mu=\left(\dot{v},\dot{x},\dot{y},\dot{u}\right)\ ,\nonumber\\
e^\mu_{(1)}&=&-\left({\dot{x}}/{(2x^2\dot{u})},1,0,0\right)\ ,\label{FrameInterpretation}\\
e^\mu_{(2)}&=&-\left({\dot{y}}/{(2x^2\dot{u})},0,1,0\right)\ ,\nonumber\\
e^\mu_{(3)}&=&-u^\mu+\left(1/{(2x^2\dot{u})},0,0,0\right)\ ,\nonumber
\end{eqnarray}
was introduced and employed  in \cite{BicPod99II}.
Mutually perpendicular spatial unit vectors $\vekte{1}$, $\vekte{2}$, $\vekte{3}$ are  orthogonal to the four-velocity
$\mathbf{u}$ of a timelike geodesic observer. Moreover, the vector $\vekte{3}$ is the projection of the principal null
direction ${\bf k}$. Equation of geodesic deviation expressed in the tetrad (\ref{FrameInterpretation}),
see \eqref{DeviationInterpretationAmplitudes} below, exhibits the transverse character of gravitational
waves with  $\vekte{3}$ representing their direction of propagation (see \cite{BicPod99II} for more details).

However, in general the tetrad (\ref{FrameInterpretation}) is {\it not} parallelly transported. Nevertheless,
orthonormal frame $\vektEbar{i}$ transported parallelly along any geodesic must exist, and at each event it
has to be related to the frame \eqref{FrameInterpretation} by a 3-dimensional rotation of its spatial part.
We may naturally assume that such relation is parametrized by the Euler angles $\varphi$, $\vartheta$, $\psi$,
\begin{equation}
\matice{c}{\vektebar{1}\crA \vektebar{2}\crA \vektebar{3}}=
\matice{ccc}{\cos{\varphi}&-\sin{\varphi}&0\crA \sin{\varphi}&\hfill\cos{\varphi}&0\crA 0&0&1}
\matice{ccc}{1&0&0\crA 0&\hfill\cos{\vartheta}&-\sin{\vartheta}\crA 0&\sin{\vartheta}&\hfill\cos{\vartheta}}
\matice{ccc}{\hfill\cos{\psi}&-\sin{\psi}&0\crA \sin{\psi}&\hfill\cos{\psi}&0\crA 0&0&1}
\matice{c}{\vekte{1}\crA \vekte{2}\crA \vekte{3}}.
\label{TransformInterpretationToParallel}\end{equation}
Now, we  impose the conditions for the parallel transport of the vectors
${\hbox{\vekt e}^{_\parallel}}_{\!(i)}(\tau)$. After straightforward but somewhat lengthy
calculation, using \eqref{Christoffel} and Eqs. \eqref{gx}-\eqref{gu},
we obtain the following system of differential equations for the Euler angles,
\begin{eqnarray}
x\,\sin\vartheta\>\dot{\varphi}&=&-\cos\psi\ ,\nonumber\\
x\,\dot{\vartheta}\,&=&\ \sin\psi\ ,\label{EquationsAngles}\\
x\sin\vartheta\>\dot{\psi}&=&-\dot{y}\sin\vartheta+\cos\vartheta\cos\psi\ .\nonumber
\end{eqnarray}
For a given timelike geodesic, characterized by specific functions $x(\tau)$, $y(\tau)$, we thus
obtain a specific rotation described by $\varphi(\tau)$, $\vartheta(\tau)$, $\psi(\tau)$ which
define the parallelly transported orthonormal frame $\vektEbar{i}$ rotating with respect
to \eqref{FrameInterpretation}.

Interestingly, the situation simplifies considerably for geodesics restricted
to ${y=0}$. The vector $e^\mu_{(2)}=\left(0,0,-1,0\right)$ is orthogonal to this
hypersurface and it is parallelly transported. It only remains to find a suitable rotation of the
vectors  ${\hbox{\vekt e}_{(1)}}$, ${\hbox{\vekt e}_{(3)}}$,
which may be obtained from \eqref{EquationsAngles} for a particular choice
${\vartheta=\pi/2}$, ${\psi=\pi}$, implying
\begin{equation}
\dot{\varphi}={x^{-1}(\tau)}\ .
\label{EquationPhi}\end{equation}
The parallelly transported orthonormal frame is thus  given by
\begin{eqnarray}
\vektebar{1}&=&-\cos\varphi\;\vekte{1}+\sin\varphi\;\vekte{3}\ ,\nonumber\\
\vektebar{2}&=&-\sin\varphi\;\vekte{1}-\cos\varphi\;\vekte{3}\ ,\label{FrameParallelHypersurface}\\
\vektebar{3}&=&-\vekte{2}\ ,\qquad \vektebar{0}=\mathbf{u}\ ,\nonumber
\end{eqnarray}
or explicitly
\begin{eqnarray}
{e^\pap}^\mu_{(1)}&=&
        \cos\varphi\left( {\dot{x}}/{(2x^2\dot{u})},1,0,0\right)-
        \sin\varphi\left(\dot{v}-{1}/{(2x^2\dot{u})},\dot{x},0,\dot{u}\right)\  ,\nonumber\\
{e^\pap}^\mu_{(2)}&=&
        \sin\varphi\left( {\dot{x}}/{(2x^2\dot{u})},1,0,0\right)+
        \cos\varphi\left(\dot{v}-{1}/{(2x^2\dot{u})},\dot{x},0,\dot{u}\right)\ ,\label{FrameParallelHypersurfaceExpl}\\
{e^\pap}^\mu_{(3)}&=&\left(0,0,1,0\right)\ ,\qquad
{e^\pap}^\mu_{(0)}=\left(\dot{v},\dot{x},0,\dot{u}\right)\ .\nonumber
\end{eqnarray}
The vectors ${\hbox{\vekt e}_{(1)}}$, ${\hbox{\vekt e}_{(3)}}$ of \eqref{FrameInterpretation}
--- the latter indicating the direction of propagation of gravitational waves ---
\emph{rotate with respect to parallelly transported frames along any timelike geodesic}, where the angle
$\varphi$ is obtained by integration of Eq. \eqref{EquationPhi}. Notice that
${\dot{\varphi}\to\infty}$ as ${x\to0}$.

\subsection{Frames parallelly transported along any null geodesic}
In order to find parallel frames along null geodesics it is more convenient to employ
the formalism of null complex tetrads
$\{{\bf k}, {\bf l}, {\bf m}, {\bar{\bf m}}\}$ such that
${{\bf k}\cdot{\bf l}=-1}$, ${{\bf m}\cdot{\bar{\bf m}}=1}$, see, e.g.,
\cite{kramerbook}. We may start with the natural tetrad \eqref{privtetrad} and
perform its boost followed by a null rotation with fixed~{\bf l},
\begin{eqnarray}
 {\bf k}' &=& B\,{\bf k}+\bar{K}\,{\bf m}+K\,{\bar{\bf m}}+K\bar{K}B^{-1}\,{\bf l}\ , \nonumber\\
 {\bf l}' &=& B^{-1}\,{\bf l}\ , \label{boost}\\
 {\bf m}' &=& {\bf m}+KB^{-1}\,{\bf l}\ ,\nonumber
\end{eqnarray}
where the parameters are given by
\begin{equation}
 K=\dot{\zeta}\ ,\quad B=\dot{v}-F/(2Q^2)\,\dot{u}\ ,\quad
 2\,\dot{\zeta}\dot{\bar\zeta}-2\,Q^2\dot{u}\dot{v}+F\,\dot{u}^2=0\ ,\label{param}
\end{equation}
cf. \eqref{metric}. We arrive at the null tetrad
\begin{eqnarray}
 {\bf k}' &=& \dot{v}\,\pa_v+\dot{\zeta}\,\pa_\zeta+\dot{\bar\zeta}\,\pa_{\bar\zeta}+\dot{u}\,\pa_u\ , \nonumber\\
 {\bf l}' &=& FQ^{-2}(2Q^2\dot{v}-F\dot{u})^{-1}\,\pa_v+2(2Q^2\dot{v}-F\dot{u})^{-1}\,\pa_u\ , \label{tetint}\\
 {\bf m}' &=& FQ^{-2}\dot{\zeta}(2Q^2\dot{v}-F\dot{u})^{-1}\,\pa_v+\pa_{\bar\zeta}+2\dot{\zeta}(2Q^2\dot{v}-F\dot{u})^{-1}\,\pa_u\ ,\nonumber
\end{eqnarray}
for which the null vector ${\bf k}'$ is obviously tangent to the corresponding null
geodesic and it is, of course, parallelly transported along this.
We wish to find a tetrad such that the remaining null vectors are
also parallel. This is achieved by performing
additional null rotation --- this time with fixed ${\bf k}'$ ---
and a spatial rotation in the transverse plane,
\begin{eqnarray}
 {\bf k}^\pap &=& {\bf k}'\ , \nonumber\\
 {\bf l}^\pap &=& {\bf l}'+\bar{L}\,{\bf m}'+L\,{\bar{\bf m}}'+L\bar{L}\,{\bf k}'\ , \label{rot}\\
 {\bf m}^\pap &=& \exp(\hbox{i}\phi)\,({\bf m}'+L\,{\bf k}')\ .\nonumber
\end{eqnarray}
We have to choose the complex function $L$ and the real function
$\phi$ of the affine parameter such that ${\bf l}^\pap$,
${\bf m}^\pap$ and ${\bar{\bf m}}^\pap$ are parallelly transported.
Calculation using \eqref{param} and the Christoffel symbols for \eqref{metric}
leads to the system of differential equations
\begin{eqnarray}
 \dot{\phi}&=&\hbox{i} \frc{\dot{\zeta}-\dot{\bar\zeta}}{Q}
    +\frc{{\dot{u}}^2}{2\dot{\zeta}\dot{\bar\zeta}}\,\hbox{i} [Q(\dot{\zeta} H_{,\zeta}-\dot{\bar\zeta} H_{,\bar\zeta})
    -(\dot{\zeta}-\dot{\bar\zeta})H]\ , \nonumber\\
 \dot{L}&=&\Big\{
 \frc{\dot{\zeta}-\dot{\bar\zeta}}{Q}
    +\frc{{\dot{u}}^2}{2\dot{\zeta}\dot{\bar\zeta}}[Q(\dot{\zeta} H_{,\zeta}-\dot{\bar\zeta} H_{,\bar\zeta})
    -(\dot{\zeta}-\dot{\bar\zeta})H]  \Big\}\,L
    -\frc{1}{Q}-\frc{{\dot{u}}^2}{2\dot{\zeta}\dot{\bar\zeta}}(QH_{,\bar\zeta}-H)
 \ .\label{PhiL}
\end{eqnarray}
In real coordinates \eqref{real} of the metric \eqref{Metric}
these conditions take the form
\begin{eqnarray}
\dot{\phi}&=&-\frc{\dot{y}}{x}-\frc{4\dot{u}^2}{\dot{x}^2+\dot{y}^2}
  \left[\,{x(\dot{x}\G_{,y}-\dot{y}\G_{,x})+\dot{y}\G}\,\right]
   \ ,\nonumber\\
\dot{L}_1-L_2\,\dot{\phi}&=& \frc{2\sqrt{2}\,\dot{u}^2}{\dot{x}^2+\dot{y}^2}\,\left({x\G_{,x}-\G}\right)-\frc{1}{\sqrt{2}\,x}
   \ ,\label{NullEquationsParameters}\\
\dot{L}_2+L_1\,\dot{\phi}&=& \frc{2\sqrt{2}\,\dot{u}^2}{\dot{x}^2+\dot{y}^2}\,\,x\G_{,y}\ ,\nonumber
\end{eqnarray}
where ${L_1, L_2}$ denote real and imaginary parts of  $L$,  ${L\equiv L_1+\hbox{i}L_2}$.
These equations are more complicated than the corresponding expressions \eqref{EquationsAngles} for parallel
frames along timelike geodesics. In particular, they explicitly depend on the structural function $\G$.

However, expressions \eqref{NullEquationsParameters} again simplify  for null geodesics  on the
hypersurface ${y=0}$ in the Kundt spacetimes  \eqref{GKvadr-q}. The simplest parallelly
transported frame is then obtained by setting
\begin{equation}
 \phi=0\ ,\qquad L_2=0\ ,\qquad
 \dot{L}_1=2\sqrt{2}\,(n-1)\,x^n{\dot{x}^{-2}}\,\dot{u}^2-(\sqrt{2}\,x)^{-1}\ .
 \label{NullFrameEquationL1Special}
\end{equation}
In real coordinates ${x, y}$, the frame is thus explicitly given  by
\begin{eqnarray}
k^{\pap\mu}&=&\left(\dot{v},\dot{x},0,\dot{u}\right)\ ,\nonumber\\
l^{\pap\mu}&=&\Big(L_1^2\dot{v}+2\dot{u}\dot{x}^{-2}(v^2+{\G}/{x})(1+\sqrt{2}L_1\dot{x})
     ,L_1^2\dot{x}+\sqrt{2}L_1,\nonumber\\
&&\hskip35mm 0,L_1^2\dot{u}+2\dot{u}\dot{x}^{-2}(1+\sqrt{2}L_1\dot{x})\Big)
   \ ,\label{NullFrameParallelHypersurface}\\
m^{\pap\mu}_1&=&\left(L_1\dot{v}+\sqrt{2}\,\dot{u}\dot{x}^{-1}(v^2+{\G}/{x}),L_1\dot{x}
    +{1/\sqrt{2}},0,L_1\dot{u}+\sqrt{2}\,\dot{u}\dot{x}^{-1}\right)
   \ ,\nonumber\\
m^{\pap\mu}_2&=&(0,0,{1/\sqrt{2}},0)\ .\nonumber
\end{eqnarray}
where ${{\bf m}^\pap ={\bf m}^\pap_1+\hbox{i}\,{\bf m}^\pap_2}$.
For particular null  geodesics of the form \eqref{AnzatzUt}  given by equations \eqref{AnzatzUt-1},
\eqref{AnzatzUt-2} with ${\epsilon=0}$, ${C=0}$ we obtain
\begin{equation}
\dot{L}_1=\textstyle{\frac{1}{\sqrt{2}}}\,(n+2a+2)\,{x^{-1}(\tau)}\ .
\label{NullFrameEquationL1Anzatz}\end{equation}
This resembles  analogous expression \eqref{EquationPhi} for timelike geodesics.

\section{Frame components of curvature tensor and geodesic deviation}
\label{project}

Now we  evaluate explicitly the components of the Riemann tensor in the above
tetrads which are parallelly transported  along timelike  or null  geodesics in
vacuum Kundt spacetimes.

\subsection{Curvature components evaluated along timelike geodesics}
Considering \eqref{Riemann}, we first project the Riemann tensor onto
the ``interpretation'' frame \eqref{FrameInterpretation},
\begin{eqnarray}
R_{(0)(1)(0)(1)}=R_{(0)(1)(1)(3)}=R_{(1)(3)(1)(3)}&=&\ \A_+ \ ,\nonumber\\
R_{(0)(2)(0)(2)}=R_{(0)(2)(2)(3)}=R_{(2)(3)(2)(3)}&=&-\A_+ \ , \label{RiemannInterpretation}\\
R_{(0)(1)(0)(2)}=R_{(0)(1)(2)(3)}=R_{(0)(2)(1)(3)}=R_{(1)(3)(2)(3)}&=&-\A_\times \ ,\nonumber
\end{eqnarray}
where the amplitudes are
\begin{equation}
\A_+=-2x\dot{u}^2{\G}_{,xx}\ ,\qquad
\A_\times=2x\dot{u}^2{\G}_{,xy}\ ,
\label{Amplitudes}\end{equation}
see also \cite{BicPod99II}. A straightforward algebraic calculation using the
transformation \eqref{TransformInterpretationToParallel} leads to the following
non-vanishing components of the curvature tensor in the orthonormal frame $\vektEbar{i}$
parallelly transported along a timelike geodesic,
\begin{eqnarray}
-R^\pap_{(2)(3)(2)(3)}&=&
     \M(\cos^2\!\varphi-\sin^2\!\varphi\cos^2\!\vartheta)+\N\cos\vartheta\sin2\varphi
     =R^\pap_{(0)(1)(0)(1)}\ ,
     \nonumber\\
-R^\pap_{(0)(1)(2)(3)}&=&
     \N(\cos^2\!\varphi-\sin^2\!\varphi\cos^2\!\vartheta)-\M\cos\vartheta\sin2\varphi\ ,
     \nonumber\\
-R^\pap_{(1)(3)(1)(3)}&=&
     \M(\sin^2\!\varphi-\cos^2\!\varphi\cos^2\!\vartheta)-\N\cos\vartheta\sin2\varphi
     =R^\pap_{(0)(2)(0)(2)}\ ,
     \nonumber\\
R^\pap_{(0)(2)(1)(3)}&=&
     \N(\sin^2\!\varphi-\cos^2\!\varphi\cos^2\!\vartheta)+\M\cos\vartheta\sin2\varphi\ ,
     \nonumber\\
-R^\pap_{(1)(2)(2)(3)}&=&
     \M{\textstyle\frac{1}{2}}\sin2\vartheta\sin\varphi-\N\sin\vartheta\cos\varphi
     =R^\pap_{(0)(1)(0)(3)}\ ,
     \nonumber\\
-R^\pap_{(0)(3)(2)(3)}&=&
     \N{\textstyle\frac{1}{2}}\sin2\vartheta\sin\varphi+\M\sin\vartheta\cos\varphi
     =-R^\pap_{(0)(1)(1)(2)}\ ,\label{RiemannParallel}\\
R^\pap_{(0)(3)(1)(3)}&=&
     \M\sin\vartheta\sin\varphi-\N{\textstyle\frac{1}{2}}\sin2\vartheta\cos\varphi
     =-R^\pap_{(0)(2)(1)(2)}\ ,
     \nonumber\\
-R^\pap_{(1)(2)(1)(3)}&=&
     \N\sin\vartheta\sin\varphi+\M{\textstyle\frac{1}{2}}\sin2\vartheta\cos\varphi
     =-R^\pap_{(0)(2)(0)(3)}\ ,
     \nonumber\\
R^\pap_{(1)(3)(2)(3)}&=&
     \M{\textstyle\frac{1}{2}}(1+\cos^2\!\vartheta)\sin2\varphi-\N\cos\vartheta\cos2\varphi
     =R^\pap_{(0)(1)(0)(2)}\ ,
     \nonumber\\
-R^\pap_{(0)(2)(2)(3)}&=&
     \N{\textstyle\frac{1}{2}}(1+\cos^2\!\vartheta)\sin2\varphi+\M\cos\vartheta\cos2\varphi
     =R^\pap_{(0)(1)(1)(3)}\ ,
     \nonumber\\
R^\pap_{(1)(2)(1)(2)}&=&\M\sin^2\!\vartheta=-R^\pap_{(0)(3)(0)(3)}\ ,
     \nonumber\\
R^\pap_{(0)(3)(1)(2)}&=&\N\sin^2\!\vartheta\ ,\nonumber
\end{eqnarray}
where the Euler angles satisfy \eqref{EquationsAngles},
$\M\equiv\A_+\cos2\psi+\A_\times\sin2\psi$,
$\N\equiv-\A_+\sin2\psi+\A_\times\cos2\psi$.

The above components of the Riemann tensor can be employed to study geodesic deviation. Let us consider
a displacement vector $Z^\mu(\tau)$ between two nearby particles moving along timelike geodesics.
We project $Z^\mu$ onto $\vektE{i}$, ${\Z{i} \equiv e^{(i)}_\mu Z^\mu}$,  and denote its absolute derivatives as
${\dot{Z}^{(i)}\equiv e^{(i)}_\mu(\mbox{D}Z^\mu/\mbox{d}\tau)}$,
${\ddot{Z}^{(i)}\equiv e^{(i)}_\mu(\mbox{D}^2Z^\mu/\mbox{d}^2\tau)}$.
The equation of geodesic deviation with respect to  \eqref{FrameInterpretation} is
${\ddot{Z}^{(i)}=-R^{(i)}_{(0)(j)(0)}Z^{(j)}}$,
i.e. using \eqref{RiemannInterpretation},
\begin{eqnarray}
\ddot{Z}^{(1)}&=&-\A_+\Z{1}+\A_\times \Z{2}\ ,\nonumber\\
\ddot{Z}^{(2)}&=&\hskip3.2mm\A_+\Z{2}+\A_\times \Z{1} ,\label{DeviationInterpretationAmplitudes}\\
\ddot{Z}^{(3)}&=&0\ ,\nonumber
\end{eqnarray}
see \cite{BicPod99II}.
The structure of the right-hand side exhibits the transverse character of gravitational waves
with two polarization modes (represented by the amplitudes ${\A_+, \A_\times}$) which propagate
in the spatial direction of ${\bf e}_{(3)}$. However, the left-hand side is complicated
because the  frame \eqref{FrameInterpretation} is  not parallelly propagated.
Nevertheless, using the relation \eqref{TransformInterpretationToParallel},  \eqref{EquationsAngles}
we can express $\ddot{Z}^{(i)}$ in terms of the derivatives $\Z{i}$ with respect to $\tau$.
The equation \eqref{DeviationInterpretationAmplitudes} thus takes the form
\begin{eqnarray}
\tdrdr{\Z{1}\!}{\tau}+\frc{2\dot{y}}{x}\tdr{\Z{2}\!}{\tau}-\frc{2}{x}\tdr{\Z{3}\!}{\tau}&=&
                 -\Big(\A_+-\frc{1}{\>x^2}-\frc{\dot{y}^2}{x^2}\Big)\Z{1}\nonumber\\
    &&   +\Big(\A_\times+\frc{\dot{y}\dot{x}}{\>x^2}-\frc{\ddot{y}}{x}\Big)\Z{2}
                 -\frc{\dot{x}}{\>x^2}\Z{3}\ ,\nonumber\\
\tdrdr{\Z{2}\!}{\tau}-\frc{2\dot{y}}{x}\tdr{\Z{1}\!}{\tau}&=&
                 \quad\Big(\A_\times-\frc{\dot{y}\dot{x}}{\>x^2}+\frc{\ddot{y}}{x}\Big)\Z{1}\ ,\label{DeviationInterpretation}\\
    &&   +\Big(\A_++\frc{\>\dot{y}^2}{\>x^2}\Big)\Z{2}-\frc{\dot{y}}{\>x^2}\Z{3}\ ,\nonumber\\
\tdrdr{\Z{3}\!}{\tau}+\frc{2}{x}\tdr{\Z{1}\!}{\tau}&=&
                 \frc{\dot{x}}{\>x^2}\Z{1}-\frc{\dot{y}}{\>x^2}\Z{2}+\frc{1}{\>x^2}\Z{3}\ .\nonumber
\end{eqnarray}

In the Kundt spacetimes given by $\G^{(n)}$ there are timelike geodesics restricted to
the privileged  hypersurface ${y=0}$, see \eqref{GKvadr-q}. In such a case we obtain
${\A_\times=0}$, ${\vartheta=\pi/2}$, ${\psi=\pi}$, and $\varphi$ solving \eqref{EquationPhi}.
The only  components \eqref{RiemannParallel} of the Riemann tensor with respect to the frame
\eqref{FrameParallelHypersurface} are thus
\begin{eqnarray}
-R^\pap_{(2)(3)(2)(3)}&=&\A_+\cos^2\varphi=\ R^\pap_{(0)(1)(0)(1)}\ ,\nonumber\\
-R^\pap_{(1)(3)(1)(3)}&=&\A_+\sin^2\varphi=\ R^\pap_{(0)(2)(0)(2)}\ ,\nonumber\\
-R^\pap_{(0)(3)(2)(3)}&=&\A_+\cos\varphi=-R^\pap_{(0)(1)(1)(2)}\ ,\nonumber\\
 R^\pap_{(0)(3)(1)(3)}&=&\A_+\sin\varphi=-R^\pap_{(0)(2)(1)(2)}\ ,\label{RiemannParallelHypersurface}\\
 R^\pap_{(1)(3)(2)(3)}&=&\A_+\sin\varphi\cos\varphi= R^\pap_{(0)(1)(0)(2)}\ ,\nonumber\\
 R^\pap_{(1)(2)(1)(2)}&=&\A_+=-R^\pap_{(0)(3)(0)(3)}\ .\nonumber
\end{eqnarray}
The equations of geodesic deviation \eqref{DeviationInterpretation} also considerably
simplify since ${\dot{y}=0=\A_\times}$.

\subsection{Curvature components evaluated along null geodesics}

We also calculate components of the Riemann tensor in the
parallelly transported null tetrad  \eqref{rot}. For vacuum spacetimes
the curvature tensor coincides with the Weyl tensor. It is convenient
to use standard transformations of the NP coefficients
$\Psi_j$, ${j=0,1,2,3,4}$, see e.g. \cite{kramerbook}. We start with the expression \eqref{psi4phi22}
with respect to the tetrad \eqref{privtetrad}, and
perform  boost $B$, null rotation $K$, followed
by  null rotation $L$ and spatial rotation $\phi$,
where these parameters are given by \eqref{param} and \eqref{PhiL},
respectively. We obtain the components,
\begin{equation}
\Psi_j^\pap=e^{(2-j)\rm{i}\phi} \,K^{4-j}\,(1+K\bar{L})^j\,\Psi\ ,\label{psiinparal}
\end{equation}
where, using \eqref{param}
\begin{eqnarray}
\Psi&\equiv&
B^{-2}\Psi_4=\frac{Q\,\dot{u}^2}{2(\dot{\zeta}\dot{\bar{\zeta}})^2}\,H_{,\zeta\zeta}
\nonumber\\
  &=&\frac{8\,x\,\dot{u}^2}{({\dot x}^2+{\dot y}^2)^2}\left(-G_{,xx}+iG_{,xy}\right)
    =4\,\frac{ \A_+ +i\A_\times}{({\dot x}^2+{\dot y}^2)^2}\ .\label{paramett}
\end{eqnarray}

Comparing \eqref{psiinparal} with \eqref{RiemannParallel} for timelike
geodesics (for which only $\A_+$ and $\A_\times$ are sufficient
to describe the behaviour of gravitational filed), investigation of the curvature tensor along
null geodesics is more complicated. One has to study not only
$\Psi$ but also various powers of $K$ and  ${(1+K\bar{L})}$.
However, there is a simplification for null
geodesic ${y=0}$ in vacuum Kundt spacetimes
with $\G^{(n)}$; we obtain ${\Psi=4\A_+/{\dot x}^4}$,
${\phi=0}$, ${K=\dot{x}/\sqrt2}$, and ${L=L_1}$ given by
\eqref{NullFrameEquationL1Special}, so that
\begin{equation}
\Psi_j^\pap= (L_1+\sqrt2/\dot{x})^j\,\A_+\ ,\quad j=0,1,2,3,4\ .\label{psiinparalyje0}
\end{equation}

\section{Character of the singularities in the Kundt spacetimes }
\label{chara}

Using the above results we will now discuss the behaviour
of the curvature tensor components with respect to frames parallelly propagated along timelike or null geodesics.
This will elucidate the character on the singularities, in
particular of the envelope singularity ${x=0}$.

It has been shown that such components of the Riemann tensor are proportional to the amplitudes
${\A_+}$, ${\A_\times}$, see expressions \eqref{RiemannParallel}
or \eqref{psiinparal} for timelike or null geodesics, respectively.
We calculate these amplitudes \eqref{Amplitudes} for
spacetimes given by the structural function \eqref{GObecna} with ${c=1}$
 on the hypersurface ${y=0}$ as
${\A_+=-2n(n-1)\,x^{n-1}\dot{u}^2}$, ${\A_\times=0}$.
The behaviour of $\A_+$ as
${x\to0}$ or ${x\to\infty}$ depends on the mutual
relation between $\dot{u}$ and the powers of $x$. This again justifies
our assumption \eqref{AnzatzUt} which enabled us in section~\ref{geodetikya}
to derive particular classes of exact or asymptotic geodesics, summarized in
tables~\ref{1Table}-\ref{Table-ExactSolutions1}. For all these
we obtain
\begin{equation}
\A_+=-2n(n-1)A^2\,x^{2a+n-1}\ .
\label{AmplitudesGan}\end{equation}

It is now easy to express explicitly the behaviour of this
function for exact geodesics denoted above as \ReseniA{n}, \ReseniB{n}, \ReseniC{n}.
Recall that only the geodesics \ReseniC{n} are timelike if $n>4$, and the behaviour of $\A_+$ is thus sufficient.
All other geodesics are null and for the discussion of  \eqref{psiinparalyje0}
it is necessary to evaluate also the function ${L_1+\sqrt2/\dot{x}}$.
In  table~\ref{Divergences} we present a summary of
the powers of $(\tau-\tau_0)$ corresponding to each quantity.

\begin{table}[h]
 \[
  \begin{array}{|c|c||c|c|c|c|} \hline
  \hbox{\,Class\,} & \epsilon
  & x
  &\,\,\A_+\,
  & L_1+\sqrt2/\dot{x}
  & \Psi_j^\pap=\left(L_1+\sqrt2/\dot{x}\right)^j\A_+
   \\ \hline\hline
\hbox{\ReseniA{n}}&0
           &\frc{2}{n+3}
           &-2
           &\frc{n+1}{n+3}
           &{\frc{(j-2)\,n+(j-6)}{n+3}}\\
  \hline
\hbox{\ReseniB{n}}&0
           &\,\frc{2(1-n)}{n^2-6n+1}\,
           &-2
           &\,\frc{n^2-4n-1}{n^2-6n+1}\,
           &\,{\frc{(j-2)\,n^2-4(j-3)\,n-(j+2)}{n^2-6n+1}}\,\\
  \hline  & -1\,\,\hbox{if}
           &
           &
           &
           &\\

\hbox{\ReseniC{n}}   &\  n>4
           &1
           &-2
           &\hbox{---}
           &\hbox{---}\\
  \hline
 \end{array}
 \]
 \caption{The powers of ${(\tau-\tau_0)}$ for various quantities which enter the components
 of the curvature tensor in frames parallelly propagated along classes of exact  geodesics
 ${y=0}$.}
\label{Divergences}
\end{table}

Interestingly, although the power-law dependence of ${x\sim(\tau-\tau_0)^p}$
on the affine parameter $\tau$ varies for different classes
\ReseniA{n}, \ReseniB{n}, and \ReseniC{n} of  geodesics, the dependence of the curvature tensor amplitude
$\A_+$ is \emph{the same}, namely
\begin{equation}
\A_+\sim (\tau-\tau_0)^{-2}\ .
\label{AAmplitudesGan}\end{equation}
This is the consequence of the fact that ${2a+n-1=2(q-1)/p+(n-1)}$, cf.
\eqref{ExactSolutionConstants}, and also ${q=-(n-1)p/2}$, which can
be seen from table~\ref{Table-ExactSolutions1}, so that ${2a+n-1=-2/p}$.
Similarly, the relation \eqref{NullFrameEquationL1Anzatz} gives
$\dot{L}_1\sim x^{-1}\sim (\tau-\tau_0)^{-p}$,
implying ${L_1\sim  (\tau-\tau_0)^{1-p}}$.
Since ${{\dot x}\sim(\tau-\tau_0)^{p-1}}$, both  ${L_1}$ and ${1/\dot x}$ exhibit the
same behaviour,
\begin{equation}
L_1+\sqrt2/\dot{x}\sim  (\tau-\tau_0)^{1-p}\ .
\label{Ladotx}\end{equation}
Let us now discuss specific behaviour of the curvature tensor in some more detail.

The only class which admits \emph{timelike geodesics} is \ReseniC{n} when ${n>4}$. This is characterized by
${a=-(n+1)/2}$ and $p=1$, i.e. ${x\sim(\tau-\tau_0)}$. The
amplitude \eqref{AAmplitudesGan}  thus  behaves as
\begin{eqnarray}
\lim_{x\rightarrow0}\A_+  &\sim&\lim_{\tau\rightarrow\tau_0}{(\tau-\tau_0)^{-2}}=\infty\ ,\nonumber\\
\lim_{x\rightarrow\infty}\A_+&\sim&\lim_{\tau\rightarrow\infty}{(\tau-\tau_0)^{-2}}=0\ .
\label{DivergencesCq}\end{eqnarray}
Obviously, all non-trivial components \eqref{RiemannParallelHypersurface} of the curvature tensor with respect to frames parallelly
transported along these timelike geodesics approaching ${x=0}$ \emph{diverge}.
The envelope singularity of the radiative Kundt spacetimes is thus a \emph{non-scalar curvature
singularity} which is reached in a finite value of the proper time. On the other hand, the spacetimes
become flat in the region ${x\to\infty}$ for these timelike geodesics.

Analogous discussion concerning null geodesics is more
complicated since the components \eqref{psiinparalyje0}
depend on $\A_+$ and also on powers of the
function ${(L_1+\sqrt2/\dot{x})}$.
The class of exact \emph{null geodesics} \ReseniA{n} is characterized by
${a=-(n+1)}$ and ${p=2/(n+3)}$, so that
${x\sim(\tau-\tau_0)^{2/(n+3)}}$,
${\A_+\sim x^{-(n+3)}\sim (\tau-\tau_0)^{-2}}$.
The relation \eqref{Ladotx} reads
$(L_1+\sqrt2/\dot{x})\sim  (\tau-\tau_0)^{(n+1)/(n+3)}$.
The frame components of the curvature tensor are thus
\begin{equation}
\Psi_j^\pap\sim (\tau-\tau_0)^{[(j-2)\,n+(j-6)]/(n+3)}\ .
\label{Psij}
\end{equation}
The lowest power appears for  ${j=0}$, namely
${\Psi_0^\pap=\A_+\sim (\tau-\tau_0)^{-2}}$, whereas the highest one for ${j=4}$,
${\Psi_4^\pap\sim (\tau-\tau_0)^{2(n-1)/(n+3)}}$.
Obviously, there is \emph{always a curvature singularity at} ${x=0}$ since the
frame component of the curvature tensor $\Psi_0^\pap$ diverges,
\begin{equation}
\lim_{x\rightarrow0}\Psi_0^\pap \sim\lim_{\tau\rightarrow\tau_0}{(\tau-\tau_0)^{-2}}=\infty\ .
\label{DivergencesAq}\end{equation}
Other components also diverge at ${x=0}$ if
the corresponding power in \eqref{Psij} is negative, i.e. for
\begin{equation}
j<2+\frc{4}{n+1}\ .
\label{singcond}
\end{equation}
Thus, $\Psi_0^\pap$, $\Psi_1^\pap$, $\Psi_2^\pap$ always diverge, $\Psi_3^\pap$ diverges at
${x=0}$ for spacetime with ${n=2}$, whereas $\Psi_4^\pap$ always approaches zero there.
For null geodesics of this class \ReseniA{n} approaching the region ${x\to\infty}$ the situation is completely the
opposite: $\Psi_0^\pap$, $\Psi_1^\pap$, $\Psi_2^\pap$ always vanish
asymptotically, whereas
\begin{equation}
\lim_{x\rightarrow\infty}\Psi_4^\pap\sim\lim_{\tau\rightarrow\infty} (\tau-\tau_0)^{2(n-1)/(n+3)}=\infty\ .
\label{DivergencesAqq}\end{equation}

Behaviour of the curvature tensor along \emph{null geodesics} \ReseniB{n} with
${a=-2n/(n-1)}$, $p=2{(1-n)}/(n^2-6n+1)$,  depends on the parameter $n$ of the specific
function \eqref{GObecna} of the spacetime. In particular, ${x\sim(\tau-\tau_0)^p}$ so that
\begin{eqnarray}
\hbox{for}\quad n<6:\
  &&x\rightarrow0\Leftrightarrow\tau\rightarrow\tau_0\ ,\qquad
    x\rightarrow\infty\Leftrightarrow\tau\rightarrow\infty\ ,\nonumber\\
\hbox{for}\quad n\geq6:\
  &&x\rightarrow0\Leftrightarrow\tau\rightarrow\infty\ ,\qquad
    x\rightarrow\infty\Leftrightarrow\tau\rightarrow\tau_0\
    .\label{condd}
\end{eqnarray}
The behaviour is thus more complicated. The amplitude ${\A_+=\Psi_0^\pap}$ is
proportional to ${(\tau-\tau_0)^{-2}}$, i.e. singular as ${\tau\to\tau_0}$,  other components
of the curvature tensor are
\begin{equation}
\Psi_j^\pap\sim (\tau-\tau_0)^{[(j-2)\,n^2-4(j-3)\,n-(j+2)]/(n^2-6n+1)}\ .
\label{PsijB}
\end{equation}
The frame component diverges as ${\tau\to\tau_0}$ if
the corresponding power of \eqref{PsijB}, which is equal to ${j(1-p)-2}$, is negative,
i.e. (considering ${p<1}$) when
\begin{equation}
j<\frc{2}{1-p}=2-\frc{4(n-1)}{n^2-4n-1}\ .
\label{singcondB}
\end{equation}
The behaviour for ${\tau\to\infty}$ is exactly the complementary one.
In particular, for spacetimes with ${n=2}$ the components $\Psi_0^\pap$, $\Psi_1^\pap$, $\Psi_2^\pap$
diverge at ${x=0}$ whereas $\Psi_3^\pap$, $\Psi_4^\pap$ approach zero there.
For ${n=3}$ the components $\Psi_0^\pap$, $\Psi_1^\pap$, $\Psi_2^\pap, \Psi_3^\pap$
diverge at ${x=0}$ while $\Psi_4^\pap$ is regular, for ${n=4}$
all the components diverge. In spacetimes ${n=5}$ with ${p=2}$
the condition reads ${j>-2}$ so that all components
$\Psi_j^\pap$ diverge as ${x\to0}$, but all approach zero  as
${x\to\infty}$. For ${n=6}$ the condition \eqref{singcondB} is
satisfied only for ${j=0}$ so that, considering \eqref{condd},
$\Psi_1^\pap, \Psi_2^\pap, \Psi_3^\pap, \Psi_4^\pap$ diverge
at ${x=0}$, and $\Psi_0^\pap$ diverges as ${x\to\infty}$. For
spacetimes with very large parameter $n$ the condition is
satisfied when $j=0$ or $j=1$ so that
$\Psi_2^\pap, \Psi_3^\pap, \Psi_4^\pap$ diverge
at ${x=0}$, and $\Psi_0^\pap, \Psi_1^\pap$ diverge as ${x\to\infty}$.

The above described behaviour of the curvature tensor near ${x=0}$ and
${x=\infty}$ has been obtained using explicit privileged classes
\ReseniA{n},  \ReseniB{n}, and  \ReseniC{n} of exact geodesics on
the hypersurface ${y=0}$. This can be confirmed by (the only)
``power-law" geodesics of the type \eqref{ExactSolution-y}
which admit ${y\not=0}$. These null geodesics in spacetimes with ${n=2}$
are explicitly given by \eqref{ExactSolution-21}. There is
$x\sim(\tau-\tau_0)^{2/5}\sim y$, $u\sim(\tau-\tau_0)^{-1/5}$ so
that, according to \eqref{Amplitudes}, ${\A_\times=0}$, ${\A_+\sim x{\dot u}^2
\sim(\tau-\tau_0)^{-2}}$, and
${K\sim  (\tau-\tau_0)^{-3/5}}$, ${(1+K\bar{L})=const}$. Consequently,
from \eqref{psiinparal}  we obtain
\begin{equation}
\Psi_j^\pap\sim (\tau-\tau_0)^{3j/5-2}\ ,\label{D2}
\end{equation}
which implies a
divergence of the components $\Psi_0^\pap, \Psi_1^\pap, \Psi_2^\pap, \Psi_3^\pap,$ at
${x=0}$. On the other hand, ${\Psi_4^\pap\sim (\tau-\tau_0)^{2/5}\sim
x}$,
which indicates a singularity at ${x\to\infty}$.

Finally, we investigate the behaviour of the curvature
tensor along more general classes of \emph{asymptotic} geodesics
approaching  the regions ${x=0}$ or ${x=\infty}$. These
have been described in previous section~\ref{ChapAnzatz-Discussion}. The explicit class
of null geodesics \ref{TableNula-1} is characterized by ${x\sim(\tau-\tau_0)^{2/(2-k)}}$
where ${k=2a+n+1}$, see tables~\ref{1Table} and  \ref{2Table}.
In view of \eqref{AmplitudesGan} we obtain
\begin{equation}
\A_+\sim x^{k-2}\sim(\tau-\tau_0)^{-2}\ ,\label{asy1}
\end{equation}
so that the component ${\Psi_0^\pap=\A_+}$ diverges at ${x=0}$ as
${\tau\to\tau_0}$. Other components are given by
${\Psi_j^\pap\sim (\tau-\tau_0)^{[(j-2)k+4]/(k-2)}}$. The same
applies to asymptotic geodesics \ref{TableNula-4} which is a
particular subcase of \ref{TableNula-1} for ${k=n-3}$.
The classes \ref{TableNula-2} and \ref{TableNula-3}  exhibit somewhat
different dependence. It turns out that
${\Psi_j^\pap\sim (\tau-\tau_0)^{(n-3+j)/2}}$ and
${\Psi_j^\pap\sim (\tau-\tau_0)^{(n-7+j)/2}}$, respectively.
Therefore, as ${x\to0}$ with ${\tau\to\tau_0}$, only the component
${\Psi_0^\pap}$ diverges for ${n=2}$  in case \ref{TableNula-2},
and for ${n=6}$  in case \ref{TableNula-3}. Other components of
the curvature tensor are regular at ${x=0}$.

Concerning the complementary asymptotics  ${x\to\infty}$, for the
case \ref{TableNekonecno-1} the functional dependence of $x$ is
the same as in the case \ref{TableNula-1} (cf. table~\ref{4Table})
so that the behaviour of the amplitude is also given by \eqref{asy1}.
With ${\tau\to\infty}$ the component ${\Psi_0^\pap}$
vanishes, but ${\Psi_j^\pap}$ diverge when
${j>2-4/k>2}$. Since \ref{TableNekonecno-2} and \ref{TableNekonecno-3} are
particular cases of \ref{TableNekonecno-1} for ${k=-1}$, all the
components ${\Psi_j^\pap}$ vanish asympotically along these
special geodesics. Finally, there are three subclasses
\ref{TableNekonecno-A}, \ref{TableNekonecno-B},
\ref{TableNekonecno-C}. For ${C=0}$, ${x\to\infty}$, they asymptotically approach
the exact geodesics \ReseniA{n}, \ReseniB{n}, \ReseniC{n} which we
have discussed in detail at the beginning of this section. For
${C\not=0}$ there are particular null geodesics of the form
${x\sim\sqrt{\tau-\tau_0}}$, for which
${\Psi_j^\pap\sim (\tau-\tau_0)^{(j-n-3)/2}}$ in case \ref{TableNekonecno-A}, and
${\Psi_j^\pap\sim (\tau-\tau_0)^{(j-7)/2}}$ in case \ref{TableNekonecno-B}.
In both these special cases, all components vanish as ${x\to\infty}$, i.e.
as ${\tau\to\infty}$.

To summarise our somewhat lengthy discussion: typical components
(like ${\A_+=\Psi_0^\pap}$)
of the Riemann tensor with respect to parallelly propagated frames {\it diverge}
as $x\rightarrow0$ (which is reached for ${\tau\to\tau_0}$) {\it both for timelike and null geodesics},
see relations \eqref{DivergencesCq}, \eqref{Psij}-\eqref{singcond}, \eqref{PsijB},
\eqref{D2}, and \eqref{asy1}.
Concerning the behaviour in the complementary region ${x\rightarrow\infty}$ of the Kundt
spacetimes there are also divergences in other components  of the curvature tensor (typically ${\Psi_4^\pap}$)
as ${\tau\to\infty}$. However, this is not so along all null geodesics
and, in particular, along timelike geodesics of the class
\ReseniC{n} for which ${\A_+\to 0}$, see \eqref{DivergencesCq}.
Further analysis is thus required to fully understand the global structure of null and timelike
infinities of these spacetimes.

\section{Conclusions}
In the present work we have studied vacuum solutions to Einstein's equations
which represent exact radiative spacetimes of the Kundt type. Our main objective
was to investigate the geodesic motion and
to elucidate the character of the singularities contained.

We found, in section~\ref{geodetikya}, particular families of exact geodesics denoted as
\ReseniA{n}, \ReseniB{n}, \ReseniC{n} (cf. table~\ref{Table-ExactSolutions1}),
and also classes of  approximate solutions to the geodesic equations
on the hypersurface ${y=0}$ as ${x\to0}$ (\ref{TableNula-1} to \ref{TableNula-3})
or ${x\to\infty}$ (\ref{TableNekonecno-1} to \ref{TableNekonecno-C}),
see tables~\ref{2Table} and~\ref{4Table}. These  asymptotic families are quite
large as they contain several independent parameters, e.g., $k$, $A$, $C$, $\tau_0$, $u_0$.

In  section~\ref{para} we presented  tetrads
that are parallelly transported along \emph{any} timelike or null geodesic.
Somewhat surprisingly, these could be given in a closed explicit form.
In particular, we demonstrated that the spatial direction  $\vekte{3}$
of propagation of the Kundt gravitational wave \emph{always rotates} with respect to
parallelly propagated frames. This is similar to analogous  behaviour
which was discovered previously in the Siklos spacetimes with ${\Lambda<0}$ \cite{Podolsky98sik}.
It can be explained by the fact that both Kundt and Siklos waves belong to a larger
class of generalized Kundt spacetimes  for which the spin coefficient $\tau$
is non-vanishing \cite{GDP04}, see also \cite{PodOrt03}.

Subsequently, in section~\ref{project} we calculated projections of the curvature tensor onto
the frames that are parallelly propagated along arbitrary timelike or null geodesic. Not only
could these components  be used for studies of geodesic deviation,
see equation \eqref{DeviationInterpretation}, but they are also crucial for discussion of
the character of singularities.

This is done in the final section~\ref{chara}. From the frame components of the curvature
tensor we explicitly demonstrated that there is a \emph{physical singularity at\,}  ${x=0}$
which is reached in a finite value of the affine parameter $\tau_0$  both for timelike and null geodesics.
Although all scalar curvature invariants of these Kundt spacetimes identically vanish,
the expanding envelope of the rotated wave surfaces ${u=const}$. (see figure~\ref{p1}) that is localized
on ${x=0}$ is a \emph{(non-scalar) curvature singularity},
according to the classification scheme introduced in \cite{EllisSchmidt}.
It is not a ``mild'' quasi-regular singularity because some of the Riemann tensor components
in the parallelly propagated orthonormal frames diverge  as $x\rightarrow0$.
It is thus not possible to extend  the Kundt solution ``smoothly'' into the inner part of the
expanding wave-front envelope ${x=0}$ by a simple flat Minkowski region.

The complementary region ${x\rightarrow\infty}$ of the Kundt spacetimes corresponds to
conformal infinity because it is reached by geodesics only  as ${\tau\to\infty}$.
There are also divergences in some components
of the curvature tensor, even if not along all  geodesics. A more detailed analysis
is still necessary to understand the global
structure of these null and timelike infinities of the Kundt wave  spacetimes.

Finally, let us observe that many of the above results obtained for
type~$N$ vacuum Kundt waves  remain valid also for \emph{conformally flat}
spacetimes \eqref{confflat} containing pure radiation. In particular, \emph{on the hypersurface} ${y=0}$ the
metric function ${\G_0=x^2+y^2}$ (including its first derivatives) is effectively identical to the function
\eqref{GObecna} with ${n=2}$, i.e.  ${\G^{(2)}=x^2-y^2}$. We thus obtain the same forms of the
geodesics and parallelly propagated frames. The tetrad components of the curvature tensor
\eqref{RiemannParallelHypersurface} are also similar, but expressions
\eqref{psiinparal} now vanish identically because the spacetimes are conformally flat
(${\Psi=0}$). However, projecting directly \eqref{Riemann} onto \eqref{NullFrameParallelHypersurface}
we obtain, for example,
$R_{\alpha\beta\gamma\delta}\, m^{\pap\alpha}_2 k^{\pap\beta} m^{\pap\gamma}_2 k^{\pap\delta}
=-2x{\dot{u}}^2=-2A^2x^{2a+1}$. For the class \ReseniA{2} of exact null geodesics one has
${a=-3}$, ${x\sim(\tau-\tau_0)^{2/5}}$, for \ReseniB{2} there is
${a=-4}$, ${x\sim(\tau-\tau_0)^{2/7}}$. In both cases the above frame component of the Riemann tensor
is proportional to ${(\tau-\tau_0)^{-2}}$ and thus diverges as ${x\to0}$ with
${\tau\to\tau_0}$. Obviously, the envelope singularity ${x=0}$ in  conformally flat
Kundt spacetimes with pure radiation is again a non-scalar curvature singularity.

\section*{Acknowledgments}

This work was supported by  the grant GACR-202/02/0735 from the Czech Republic and
the grant GAUK from the Charles University in Prague. The authors
are grateful to M.~Ortaggio and J.~B.~Griffiths for useful
comments.

\end{document}